\documentclass[11pt, letter]{article}
\usepackage{t1enc}
\usepackage[latin2]{inputenc}
\usepackage[english]{babel}
\usepackage{fullpage}
\usepackage{caption, subcaption}
\usepackage{graphicx, float}
\usepackage[toc,page]{appendix}
\usepackage{verbatim}
\usepackage{cite}
\usepackage{url}
\usepackage[normalem]{ulem}
\usepackage{subfiles}
\usepackage{xcolor}
\usepackage{amsmath} 
\usepackage{amsfonts}
\usepackage{amssymb} 
\usepackage{mathrsfs}
\usepackage{mathtools}
\usepackage{hyperref}
\usepackage{booktabs}
\usepackage{authblk}    
\usepackage{orcidlink}  

\usepackage[noabbrev]{cleveref}
\usepackage[numbers,sort&compress]{natbib}
\usepackage{listings} 
\usepackage{xcolor}   
\usepackage{cleveref}

\begin{document}

\title{Parameter estimation of gravitational-wave signals with frequency-dependent antenna responses and higher modes}

\author[1]{Pratyusava Baral\,\orcidlink{0000-0001-6308-211X}\thanks{\href{mailto:pbaral@uwm.edu}{pbaral@uwm.edu}}}
\author[2]{Soichiro Morisaki\,\orcidlink{0000-0002-8445-6747}}
\author[3]{Ish Gupta}
\author[1]{Jolien Creighton\,\orcidlink{0000-0003-3600-2406}}
\affil[1]{University of Wisconsin-Milwaukee, Milwaukee, WI 53201, USA}
\affil[2]{Institute for Cosmic Ray Research, The University of Tokyo, 5-1-5 Kashiwanoha, Kashiwa, Chiba 277-8582, Japan}
\affil[3]{Institute for Gravitation and the Cosmos, Department of Physics,
Pennsylvania State University, University Park, PA 16802, USA}
\maketitle
\begin{abstract}
We implement frequency-dependent antenna responses and develop likelihood classes (standard likelihood, multibanded likelihood, and the relative binning (RB) likelihood) capable of handling the same within the framework of \texttt{Bilby}. We validate the approximate likelihoods by comparing them with the exact likelihood for a GW170817-like signal (luminosity distance scaled such that signal-to-noise ratio $\sim$ 1000) containing higher-order modes of radiation. We use the RB likelihood to perform parameter estimation (PE) for a GW170817-like signal, including Earth-rotation effects, detector-size effects, and higher-order modes. \textcolor{black}{To the best of our knowledge, we are the first to perform proof-of-concept Bayesian parameter estimation, including all the above-mentioned effects.} We study the system in several detector networks consisting of a single 40 km Cosmic Explorer, a 20 km CE, and a present-generation detector at A+ sensitivity. The PE runs with RB take around a day to complete on a typical cluster. \textcolor{black}{We find that higher-order modes significantly improve sky localization in scenarios where only a single next-generation detector is operating. This advantage diminishes when a second detector is added to the network; however, in such cases, we observe improved recovery of the polarization angle. While ignoring higher-order modes leads to a reduction in precision, it does not introduce bias in parameter recovery. In contrast, neglecting the effects of Earth's rotation and the finite size of long-arm detectors can result in biased parameter estimates.}
\end{abstract}

\section{Introduction}
The discovery of gravitational waves (GWs) has enabled the study of compact binary coalescences (CBCs) comprising black holes and neutron stars \cite{LVKO3, 4OGC, PhysRevLett.119.161101, Abbott_2020, Abbott_2021}. The LIGO-Virgo-KAGRA (LVK) \cite{Aasi_2015, Acernese_2015, https://doi.org/10.48550/arxiv.2008.02921} collaboration has confidently detected around 90 compact CBCs with 250 more candidates from the present observing run \cite{GraceDB}. These detections have tested our understanding of gravity, cosmology, and astrophysics \citep{P1,P2, P3}. The proposed next-generation ground-based detectors, such as the Cosmic Explorer (CE) \cite{Reitze2019Cosmic} and Einstein Telescope (ET) \cite{Punturo_2010}, promise an exciting era of gravitational wave detections, offering sensitivities that extend well beyond current capabilities. The next-generation observatories are sensitive enough to detect signals from sources at cosmological distances and provide precise measurements of their parameters, playing a pivotal role in advancing our understanding of fundamental physics, astrophysics, and cosmology \cite{Maggiore_2020, evans2023, Branchesi_2023}.

To fully understand the potential of these next-generation detectors, we require estimates of how accurately various parameters can be measured. Using conventional Bayesian parameter estimation (PE) is challenging, given the large number of detections ($\mathcal{O} \sim 10^5 - 10^6$) \cite{PhysRevD.100.064060} per year, some with signal-to-noise ratios (SNRs) greater than 1000. Moreover, signals are long and we need to incorporate corrections due to Earth's rotation at low frequencies and detector-size at high frequencies \citep{malik2008, malik2009}. Ignoring these effects is known to cause biases at high SNRs \cite{PhysRevD.108.043010}. Higher-order modes of radiation become non-negligible at higher SNRs, making Bayesian PE more complex and computationally expensive. 

Fisher information matrices offer a computationally cheap way to estimate the uncertainties in measured parameters. Effects on localization due to Earth's rotation have been extensively studied in this framework \cite{PhysRevD.97.123014, PhysRevD.97.064031}. Several constraints on cosmological parameters have also been given using Fisher matrices \cite{Jin_2020}. This method has also been used for parameter estimation of intermediate mass black holes in a network of detectors and the results agree with Bayesian parameter estimation to an order of magnitude in the parameter space of interest \cite{PhysRevD.110.103002}. However, no general Fisher matrix approach taking into account all the effects due to Earth's rotation, the detector size, and higher modes exists in the literature. Also, this approximation becomes \textcolor{black}{invalid} if the parameter space has multimodalities. 

Bayesian parameter estimation does not suffer from the shortcomings of Fisher information matrices but is computationally expensive. A recent paper by Baker et al. \cite{baker2025} highlights the computational challenges of inference in next-generation detectors. Nearly all Bayesian PE in next-generation detectors involves approximating the likelihood \textcolor{black}{due to computational constraints \cite{hu2025}}.  Pre-merger localization of compact-binary sources in next-generation networks \cite{Nitz2021} has been studied using relative binning \cite{cornish2013, zackay2018relativebinningfastlikelihood}. Reduced-order models that take into account only the amplitude modulations due to Earth rotation using BNS signals lasting 90 minutes in-band from 5 Hz to 2048 Hz have been constructed for a network of detectors \cite{PhysRevLett.127.081102}. \textcolor{black}{Bayesian PE with Earth's rotation effects and detector-size effects for long BNS signals have been implemented by Baral et al. \cite{PhysRevD.108.043010} in the framework of the multibanding approximation \cite{PhysRevD.104.044062}}. Bayesian PE with relative binning has also been used to measure tidal deformabilities in ET \cite{PhysRevD.108.023018}. Similar work has been done for Laser Interferometer Space Antenna (LISA) \citep{PhysRevD.103.083011}.

In this paper, we develop a generalized framework using \texttt{Bilby} \cite{bilby_paper} to generate waveforms that include effects due to Earth rotation, the finite size of detectors, and higher modes. We develop four new likelihood classes: the exact likelihood including all effects; the multibanding likelihood capable of handling all effects; a relative binning likelihood capable of handling the dominant mode of radiation; a mode-by-mode relative binning likelihood capable of handling all modes of radiation. We do illustrative PE for two GW170817-like \cite{PhysRevLett.119.161101} sources in four network configurations: a single 40 km CE, a network of LIGO-India and a 40 km CE; a network of a 40 km long and a 20 km long CE; and a network comprising all detectors. The luminosity distance has been scaled to fix the SNR to $\sim$ 1000. the  We study the roles played by the higher-order multiples of radiation in these networks. 

The rest of the paper is organized as follows. Section \ref{methods} describes the methods implemented for this study. Section \ref{simulations} details the gravitational wave sources, waveform families, and detection networks used in our analysis. Finally, Section \ref{results} presents the validation of our methods and the main results of the study.

\section{Methods}\label{methods}
\subsection{Frequency-dependent antenna responses} 
In general relativity, the metric perturbation $h_{ij}$ corresponding to the GW in the transverse traceless gauge is characterized by two polarization components $h_+$ and $h_\times$ as described by \cite{Creighton2011},
\begin{equation} 
    h_{ij} = h_+ e^+_{ij} + h_\times e^\times_{ij}
\end{equation}
The GW polarization tensors ($e^+_{ij}, e^\times_{ij}$) can be expressed in terms of basis vectors in the transverse plane \textcolor{black}{($e^a_i$)} given by,
\begin{eqnarray}
e^+_{ij} := e^1_i e^1_j - e^2_i e^2_j \\
 e^\times_{ij} := e^1_i e^2_j +e^2_i e^1_j
\end{eqnarray}
Here $i = x,y$ runs over the spatial indices on the transverse plane with the z-axis being the normal vector from the detector to the source, and $a = 1,2$ denote the arms of the interferometer. Thus, loud sources in next-generation detectors observed from a frame co-rotating with the Earth (the rest-frame of ground-based detectors) shift in the sky, as the signal evolves. The Earth's rotation also introduces an integrated Doppler shift. This is because the time delay between the detector and the geocenter is different for various frequencies corresponding to the CBC signal, since lower frequencies arrive at earlier times. The effect on the beam patterns and the Doppler effect combined is called the Earth-rotation effects in this paper.

The detected GW strain as a function of time ($t$) can be expressed as $d(t) = h_D(t) + n(t)$, where $n$ represents noise and $h_D$ is the GW signal at the detector, obtained by the product of the detector tensor ($D^{ij}$) \cite{PhysRevD.96.084004} and the metric perturbations. \textcolor{black}{The detector tensor is given by,
\begin{equation}
     D_{ij}(t, f_L; \xi, I) := \frac{1}{2}\Big[D_1(t, f_L; \xi, I) e^1_i (t; \xi, I) e^1_j (t; \xi, I) - D_2(t, f_L; \xi, I) e^2_i (t; \xi, I) e^2_j (t; \xi, I)\Big]
\end{equation}
where the detector scalar ($D_a$) \cite{malik2008, malik2009} is given by
\begin{equation}\label{equation: ds}
\begin{aligned}
D_a(t, f_L; \xi, I) &:=\exp\{{\pi i f_L(1-\textbf{z} \cdot \textbf{e}^a)}\} \text{sinc}\{\pi f_L(1+\textbf{z} \cdot \textbf{e}^a)\}\\
&+ \exp\{{-\pi i f_L(1+\textbf{z} \cdot \textbf{e}^a)}\} \text{sinc}\{\pi f_L(1-\textbf{z} \cdot \textbf{e}^a)\}
\end{aligned}
\end{equation}
}
Here $f$ denotes the frequency, $I$ is the position and orientation of the interferometer \textcolor{black}{$f_L:=fL/c$ and $\text{sinc } x := \sin x / x$}. The angle uniquely defining $e^a_i$ in the transverse plane is the polarization angle ($\psi$). The right ascension ($\rm RA$), declination ($\rm dec$), $\psi$ is collectively denoted by $\xi:=({\rm RA}, {\rm dec}, \psi)$.

In terms of the polarization components, the strain at the detector is expressed as,
\begin{equation} \label{eq:h}
    h_D(t;\Theta) = \sum_{i=+,\times}F_i (t, f_L; \xi, t_c, I) h_i(t-t_c; \theta, D_L, \theta_{JN}, \phi_c) 
\end{equation}
where 
\begin{eqnarray} \label{eq:antres}
    F_+ (t, f_L; \xi, I):=D^{ij}(t, f_L; \xi, I) e^+_{ij}(t; \xi, I)\\
    F_\times (t, f_L; \xi, I):=D^{ij}(t, f_L; \xi, I) e^\times_{ij}(t; \xi, I)
\end{eqnarray}
Here $t_c$ is the arrival time of the frequency corresponding to the merger at the geocenter. The waveform polarizations ($h_+, h_\times$) depend on the intrinsic source parameters like chirp mass ($\mathcal{M}$), mass ratio ($q$) and aligned-spin parameters ($\chi_1, \chi_2$), collectively denoted by $\theta:=({\mathcal{M}, q, \chi_1 \chi_2})$, the luminosity distance $D_L$, the inclination angle $\theta_{\rm JN}$ and the coalescence phase $\phi_c$.  Collectively, these parameters are represented as $\Theta:=(\xi, \theta, D_L, \theta_{\rm JN}, \phi_c, t_c)$.

The signals are short-lived for current-generation ground-based detectors and so $t \simeq t_c$, where $t_c$ is the arrival time of the GW at the detector. Thus, the polarization tensor does not vary with time. The term $f_L \rightarrow 0 \implies D_{ij} \rightarrow 1$ (see equation \ref{equation: ds}) for present detectors as the wavelength of the GW is long compared to the size of the detector. The long-wavelength approximation makes the detector tensor a constant tensor with no dependence on frequency or the direction of the source. Thus, for the present generation detectors,
\begin{eqnarray}
    F_+ (t_c, 0; \xi, I):= e^+_{ij}(t_c; \xi, I)\\
    F_\times (t_c, 0; \xi, I):= e^\times_{ij}(t_c; \xi, I)
\end{eqnarray}
Thus, for a given pair of source and detector, the antenna-response functions $F_+$ and $F_{\times}$ are constants with no frequency dependence.

The long-wavelength approximation breaks down for a 40 km long detector at high frequencies. For finite wavelengths, the measured strain depends not only on when wavefronts that pass through the vertex of the interferometer, but also when it passes through the end of the arms \cite{malik2008, malik2009}. This introduces an additional dependence on the direction of the wave propagation, which in turn depends on the (time-changing) wavelength (or frequency). This effect is referred to as the detector-size effect in this paper.

Thus, for \textcolor{black}{next}-generation detectors we need the complete time-frequency dependence as in equation \ref{eq:antres}, with the $D^{ij}$ having additional frequency and directional dependence, $e^{+,\times}_{ij}$ having temporal dependence and an integrated Doppler shift. Ignoring these effects may introduce systematic biases in parameter estimation (PE) \cite{PhysRevD.108.043010}.

Parameter estimation is often performed in the frequency domain, necessitating the Fourier transformation of equation \ref{eq:h} to the frequency domain. Analytical computation of the Fourier transform is challenging without approximations. \textcolor{black}{We first note that for a particular quasicircular inspiral mode of a GW from a CBC} time ($\tau:=t-t_c$) and frequency are not independent parameters and are related by an invertible function. We assume that $h_i$'s oscillate rapidly compared to the antenna responses (the stationary phase approximation).  We can therefore, take the Fourier transform of the waveform polarizations and convert the times in the antenna response to frequency using \textcolor{black}{ a frequency to $\tau$ converter ($\tau(m,f)$) \footnote{In PN theory the time-to-merger for an azimuthal mode $m$ from a frequency $f$, is equivalent to the time to merger of a $m=2$ mode from a frequency $2f/m$ \cite{bustillo2015comparisonsubdominantgravitationalwave, PhysRevD.102.064002}} correct up to 2 PN for an arbitrary azimuthal mode as described in \cite{PhysRevD.52.848}} While this method is approximate, it provides sufficient accuracy for practical applications. Details of this method can be found in \cite{PhysRevD.108.043010}.

The GW polarizations are often expressed in terms of spin-weighted spherical harmonics.
\begin{equation} \label{eq:waveform}
    h_D(f; \Theta) = \sum_{i=+,\times} \sum_{|m|=0}^{\infty} F_i(t(\tau(m, f),t_c), f_L; \xi, t_c, I)\sum_{\substack{ {l=2} \\ {l>m}}}^{\infty} \Tilde{h}^{lm}_i(f; \theta, D_L, \theta_{\rm JN}, \phi_c, t_c)
\end{equation}
where \textcolor{black}{$\Tilde{h}^{lm}_+$ and $\Tilde{h}^{lm}_\times$ are the fourier transforms of }the real-valued $h^{lm}_+$ and $h^{lm}_\times$, \textcolor{black}{which} are related to the complex valued $h_{lm}(t)$ and the spin-2 weighted spherical harmonics $_{-2}Y_{lm}$ as
\begin{equation}
h^{lm}_+(t; \theta, D_L, \theta_{\rm JN}, \phi_c, t_c) - i h^{lm}_\times(t; \theta, D_L, \theta_{\rm JN}, \phi_c, t_c) = _{-2}Y_{lm}(\theta_{\rm JN}, \phi_c) h_{lm}(t; \theta, D_L, t_c)
\end{equation}

We implement the classes and functions required to generate a waveform with higher modes and frequency-dependent antenna response in a forked version of \texttt{Bilby}. To summarize, this involves creating a new frequency domain source model that creates a dictionary of the plus and cross polarizations for every mode of radiation, \textcolor{black}{functions capable of generating frequency-dependent antenna responses for any mode}. For computational efficiency, modes with the same azimuthal number $m$ are grouped, as they share the same antenna response. An exception is made for relative binning, where all modes are processed individually, as discussed in later sections.
\subsection{Sampling the Posterior}
Bayesian parameter estimation depends on calculating the posterior probability distribution $p(\Theta|d)$ from the data $d$.
\begin{equation}
    p(\Theta|d):=\frac{\mathcal{L}(\Theta)p(\Theta)}{\mathcal{Z}}
\end{equation}
The prior denoted by $p(\Theta)$ is predetermined, and the evidence $\mathcal{Z}$, the integral of the numerator over all parameters ($\Theta$), is a normalization constant. The likelihood \textcolor{black}{($\mathcal{L}(\Theta)$)} assuming that noise is stationary and gaussian is given by \textcolor{black}{$\mathcal{L}(\Theta) := p(d|\Theta) \propto \exp (- \frac{1}{2}{\rm\big<} d-h|d-h{\rm\big>}) \propto \exp ({\rm\big<}h|d{\rm\big>} - {\rm\big<}h|h{\rm\big>/2})$} where ${\rm\big<}a|b{\rm\big>} = 4 ~ \text{Re} \int_0^{\infty} \frac{a^*(f)b(f)}{S_n(f)}df$ and $S_n(f)$ is the one-sided power spectral density. 

We consider a frequency band of 5 Hz to 4096 Hz, and so we set the sampling rate to 8192 Hz. For a GW170817-like signal, the dominant mode of radiation $h_{(+,\times)}^{2,2}$ lasts in-band for approximately 2 hours, requiring waveform evaluations at $10^7$ points. We need to calculate the waveform for each azimuthal mode \texttt{m}, which takes around 10s \textcolor{black}{(assuming \texttt{IMRPhenomXPHM} waveform model)} on a single CPU. Computing the waveform for three azimuthal modes (\texttt{m=2,3,4}) takes approximately 30 seconds on a single CPU. Sampling the likelihood using a nested sampler like \texttt{dynesty} requires around $10^8$ waveform evaluations for high SNRs and is computationally prohibitive. To accelerate likelihood evaluations, we employ two approximations: multibanding and relative binning, described in subsequent sections.
\subsubsection{The Multibanding Approximation}
The multibanding approximation \cite{Cannon_2012, Vinciguerra_2017, Adams_2016, PhysRevD.104.044062} is a form of adaptive sampling that divides the entire frequency range into overlapping frequency bands where the start and end frequencies depend on the sequence of durations controlled by an accuracy factor. The duration should be computed using the highest azimuthal mode \textcolor{black}{because, at a fixed frequency, the time to merger for a higher azimuthal mode is longer.} This approximation was used to do parameter estimation of signals in Cosmic Explorer containing the dominant $h_{(+,\times)}^{2,2}$ mode \cite{PhysRevD.108.043010}. However, unlike \cite{PhysRevD.108.043010}, we compute the antenna response at every banded frequency point instead of every 4 seconds, making the algorithm more accurate without trading off performance.

We develop a new source model that can generate waveform polarizations for every azimuthal mode given an arbitrary frequency array \textcolor{black}{which is necessary for our multibanding implementation. } The highest azimuthal mode and a reference chirp mass value is used to calculate the banding times. A  tunable parameter controls the number of bands and hence the accuracy of the approximation (Referred to $L$ in \cite{PhysRevD.104.044062}).  A value of 5 results in log-likelihood errors less than unity (see figure \ref{fig:comparison}). This approximation, due to adaptive frequency binning, needs only $10^5$ evaluations per waveform, resulting in a speed-up of $\mathcal{O}$($10^2$).  
\subsubsection{The Relative Binning Approximation}
The relative binning approximation accelerates parameter estimation by leveraging the smooth variation of waveform ratios as a function of frequency across neighboring points in the parameter space in a typical set of posterior distribution. The waveform corresponding to the maximum likelihood estimate is selected as the fiducial waveform, serving as the reference for constructing approximations. Waveforms at nearby points in the parameter space are computed by applying piecewise linear modifications to the fiducial waveform, significantly reducing the computational overhead of waveform generation. 
\subsubsection{Mode-by-mode Relative Binning}
For systems that include significant contributions from higher-order modes (like mass-asymmetric systems) or precessional effects, the oscillatory nature of the frequency-domain waveform can challenge the accuracy of standard piecewise-linear approximations. While \cite{Krishna:2023bug} demonstrates that the relative binning technique remains effective even for asymmetric systems, a more robust method---mode-by-mode relative binning---was introduced by \cite{Leslie:2021ssu}, \cite{narola2023} to address the complexities arising from the presence of higher modes. This approach decomposes the waveform into its spherical harmonic modes, $(l, m)$, and applies the relative binning approximation to each mode individually. The full waveform is then reconstructed by summing the approximated contributions from all modes. This refinement improves the accuracy of waveform modeling for systems with significant higher mode contributions or precessional dynamics, while retaining the computational advantages of the relative binning framework \cite{Gupta:2024bqn}. This approach achieves required accuracy with $10^4$ evaluations per waveform, resulting in speed-ups of up to $\mathcal{O}(10^3)$ \cite{Zackay:2018qdy,Krishna:2023bug}.
\section{Simulations}\label{simulations}
For illustrative purposes, we consider  GW170817-like signals \textcolor{black}{at an optimal SNR of 1000} which we shall refer to as GW1 and GW2. The difference between $\Theta_{\rm GW1}$ and $\Theta_{\rm GW2}$ is the inclination angle. GW2 has an inclination of $\pi/2$ (commonly referred to as an edge-on configuration) and accumulates more SNR for higher azimuthal mode as seen in Table \ref{table:snr}. \textcolor{black}{The luminosity distance has been scaled so that the optimal SNR is 1000. The duration of the signal is 107 minutes, assuming the signal enters the CE band at 5Hz.} The injected parameters and the priors used for sampling are in Table \ref{table:parameters_with_priors}. 
\begin{table}[h!]
    \centering
    \renewcommand{\arraystretch}{1.5} 
    \setlength{\tabcolsep}{10pt} 
    \begin{tabular}{ccc}
        \hline
        \textbf{Parameter} & \textbf{Injected Value} & \textbf{Priors} \\ 
        \hline
        Chirp Mass ($\mathcal{M}_z$) & 1.22052 (GW1); 1.20983 (GW2) \( M_{\odot} \) & $\mathcal{M}^{\textrm{inj}}_z$ + Uniform($-10^{-5}, 10^{-5}$) \( M_{\odot} \) \\
        Mass Ratio ($q$) & 0.918 & Uniform(0.2, 1) \\
        $\chi_1^z$ & 0 & Uniform(-0.05, 0.05) \\
        $\chi_2^z$ & 0 & Uniform(-0.05, 0.05) \\
        Right Asc. (RA) & 3.44616 & Uniform(0, \( 2\pi \)) \\
        Declination (Dec) & -0.408084 & Cosine($-\pi/2$, $\pi/2$) \\
        Incl. Angle ($\theta_{\textrm{JN}}$) & 0.53 (GW1); $\pi/2$ (GW2) & Sine(0, $\pi/2$) \\
        Pol. Angle ($\psi$) & 2.212 & Uniform(0, \( \pi \)) \\
        Phase ($\phi_c$) & 5.180 & Uniform(0, \( 2\pi \)) \\
        Time at CE ($t_{CE}$) & 1187008882.45 s & $t_{CE}$ + Uniform(-0.003, 0.003)s \\
        Lum. Distance ($D_L$) & 86.35 Mpc (GW1); 46.95 Mpc (GW2) & Uniform in Vol.(10, 100) Mpc \\
        \hline
    \end{tabular}
    \caption{Injected parameters and the prior distribution of parameters. Note that the distance prior is a power law with index 2, so it ignores cosmological effects. For the dominant (2,2) mode we marginalize over phase. The time is always defined with respect to a 40km CE detector at the LIGO-Hanford site because that helps in sampling when we have only one detector.}
    \label{table:parameters_with_priors}
\end{table}

We use 4 detector configurations for our study namely:
\begin{itemize}
\item{CE}: A 40 km CE at the LIGO-Hanford site. We use the displacement PSD divided by the arm-length so that it does not include corrections due to detector size at high frequencies \cite{Srivastava_2022}. The high-frequency corrections are included in the antenna response. From henceforth CE refers to the 40 km detector.
\item{CECE20}: An additional 20 km CE is placed at the LIGO-Livingston site to form a 2-detector network.
\item{CEA1}: A 2-detector network comprising a CE at the LIGO-Hanford site and LIGO India, Aundha site at A+ sensitivity.
\item{CECE20A1}: A 3-detector network comprising a CE at the LIGO-Hanford site, a CE20 at the LIGO-Livingston site, and a LIGO India Aundha at Aplus sensitivity.
\end{itemize}

We use \texttt{IMRPhenomXPHM} \cite{PhysRevD.103.104056} to simulate the GW waveforms containing (2, 2), (3, 2), (3, 3) and (4, 4) which are injected into a "zero-noise" realization of Gaussian noise  \cite{PhysRevD.98.084016} at the detectors. \texttt{IMRPhenomXPHM} is a black hole waveform class, so the tidal deformability parameters are zero by default.  We also use aligned spin priors for our analyses. For each detector network we perform 3 inferences: including only the $m=2$ mode; excluding the (4, 4) mode alone, and including all modes. The optimal SNR due to each mode is in Table \ref{table:snr}.

\begin{table}[h!]
\centering
    \renewcommand{\arraystretch}{1.5} 
    \setlength{\tabcolsep}{10pt} 
\begin{tabular}{c|ccc|ccc}
\hline
Mode  & \multicolumn{3}{c|}{GW1}                                      & \multicolumn{3}{c}{GW2} \\ 
\multicolumn{1}{c|}{} & \multicolumn{1}{c}{CE} & \multicolumn{1}{c}{CE20} & A1 & \multicolumn{1}{c}{CE} & \multicolumn{1}{c}{CE20} & \multicolumn{1}{c}{A1} \\ \hline
(2,2) & \multicolumn{1}{c}{1000} & \multicolumn{1}{c}{503} & 41.6 & 1003   & 536   & 28.9  \\ 
(3,2) & \multicolumn{1}{c}{0.85}    & \multicolumn{1}{c}{0.46}   & 0.07  & 2.96     & 1.78     & 0.16   \\  
(3,3) & \multicolumn{1}{c}{3.21}    & \multicolumn{1}{c}{1.70}   & 0.18  & 6.37     & 3.57     & 0.26   \\ 
(4,4) & \multicolumn{1}{c}{1.47}    & \multicolumn{1}{c}{0.82}   & 0.11  & 5.72     & 3.42     & 0.31   \\ \hline
\end{tabular}
\caption{Optimal SNR collected by each mode of radiation for GW1 and GW2 in the band 5 Hz - 4096 Hz for a 40 km CE, a 20 km CE and LIGO-India.}
\label{table:snr}
\end{table}

\section{Results}\label{results}
We use mode-by-mode relative binning to approximate the likelihood for its computational efficiency. For sampling the 11-dimensional parameter space \footnote{For $m=2$ mode only run, we marginalize the phase analytically and so it is a 10 parameter run in that case.}, we utilize \texttt{dynesty}, as implemented in \texttt{Bilby}. This dynamic nested sampling algorithm is particularly well-suited for Bayesian inference in complex parameter spaces. The sampling process generates new samples using the \texttt{acceptance-walk} method. In this approach, all Markov Chain Monte Carlo (MCMC) chains run for the same duration at each iteration, with the chain length dynamically adjusted to maintain a target acceptance rate. The termination criterion for sampling is set such that the relative change in evidence remains below 0.01.
\subsection{Validation}
\begin{figure}[h!] 
    \centering
    \includegraphics[width=0.8\textwidth]{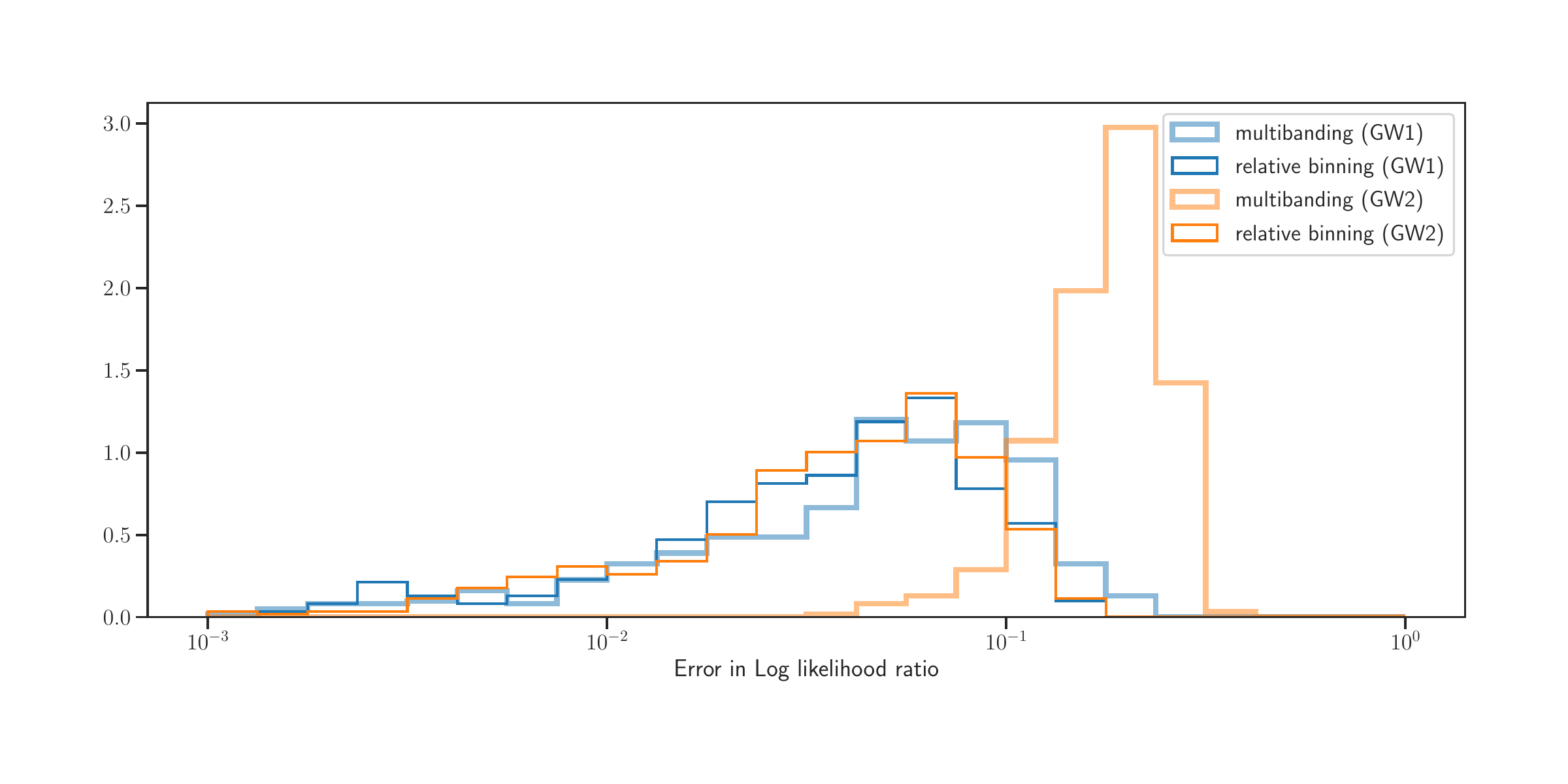} 
    \caption{The histogram of errors in log-likelihood ratio obtained using 500 samples drawn from the relative binning posterior of GW1 and GW2 in a 40 km CE. For multibanding, we use an \texttt{accuracy\_factor} of 1 and enabled linear interpolation. For relative binning, we set \texttt{chi} to 10 and \texttt{epsilon} to 0.1. It is worth noting that the histogram can shift left and right on the horizontal scale, which depends on the choice of approximation parameter. We only require that the errors are less than unity for unbiased PE.}
    \label{fig:comparison}
\end{figure}
We evaluate the accuracy of the methods described in the previous section by selecting 500 points from the posterior distributions of GW1 and GW2, observed in a 40 km CE. For each point, the exact log-likelihood ratio is calculated using the exact likelihood. As CE is the most sensitive detector considered in this study, the differences in log-likelihood ratios are expected to be even smaller for other detectors. The chosen signal has a high SNR and includes higher-order modes, making it well-suited for this analysis. Additionally, we also validate the multibanding approximation. \textcolor{black}{We tune the adjustable parameters for the approximate likelihoods such that }in both cases, the difference in log-likelihood ratios is found to be less than unity \textcolor{black}{which satisfies the criterion in \cite{Vinciguerra_2017}}, as illustrated in Figure \ref{fig:comparison}, confirming the robustness of our results. In Figure \ref{fig:comparison}, relative binning appears more accurate than multibanding; however, this is an artifact of the chosen accuracy-controlling parameters. The parameters chosen for multibanding perform worse when higher modes are present. The exact reason for this is left for future investigation. The current scheme to determine frequency bins in relative binning is solely based on the Post-Newtonian inspiral phase, but it can be extended to incorporate the effects of Earth's rotation and finite detector sizes, because their oscillatory scales in the frequency domain can be easily estimated. Frequency bins can be constructed so that the interpolation accurately captures those effects. 

\subsection{Inference using 1 Cosmic Explorer}
\begin{figure}[h!]
    \centering
    \begin{subfigure}{0.49\textwidth}
        \centering
    \includegraphics[width=\textwidth]{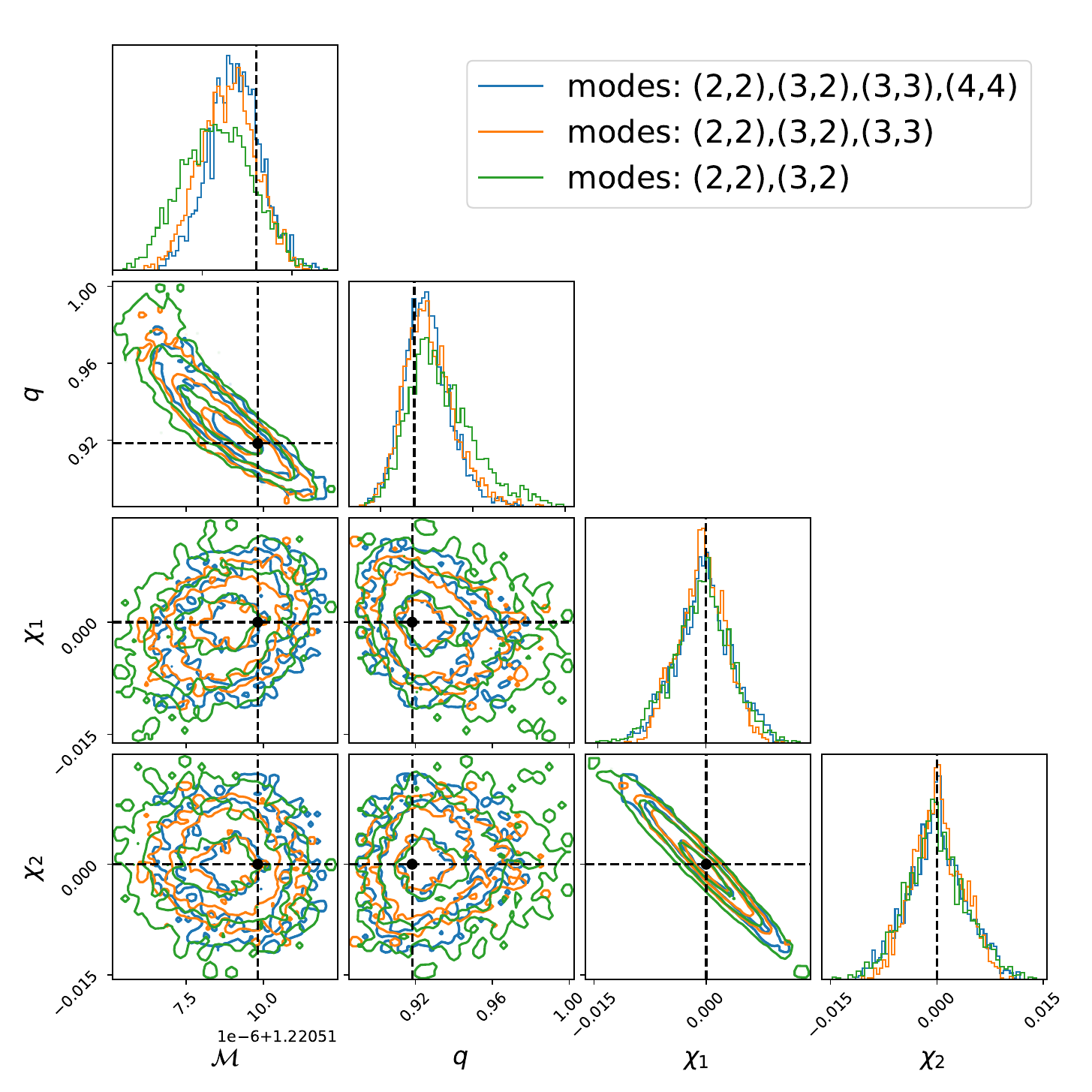}
        \caption{GW1}
        \label{fig:ce_int2}
    \end{subfigure}
    \hfill 
    \begin{subfigure}{0.49\textwidth}
        \centering
        \includegraphics[width=\textwidth]{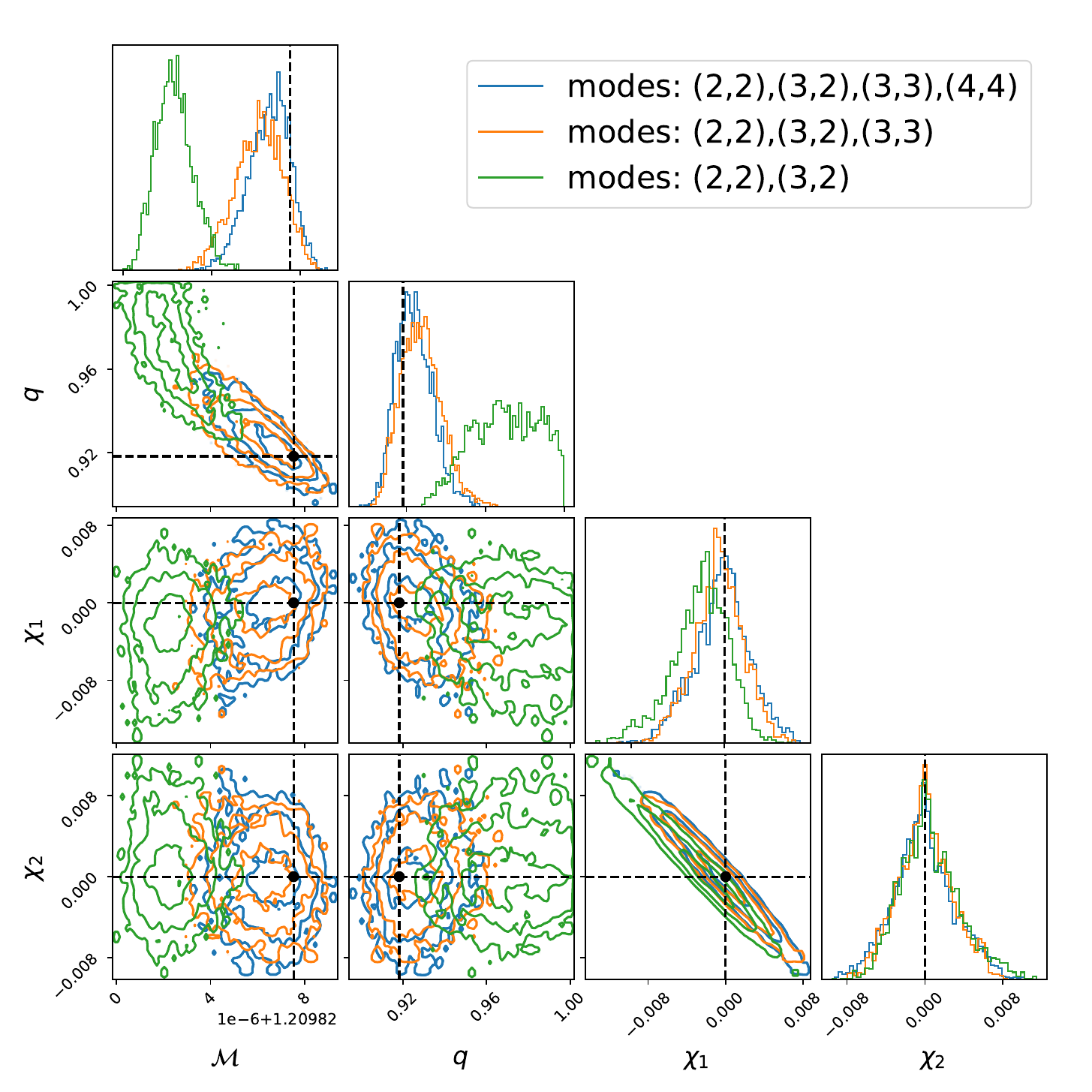}
        \caption{GW2}
        \label{fig:ce_int1}
    \end{subfigure}
    \caption{Inferred posteriors of intrinsic parameters of GW1 and GW2 in a 40km CE. The green posteriors use all modes for analysis, the orange posteriors ignore modes with \texttt{m = 3} and the blue ones ignore \texttt{m = 3} and \texttt{m = 4} modes.}
    \label{fig:side_by_side}
\end{figure}

\begin{figure}[h!]
    \centering
    \begin{subfigure}{0.49\textwidth}
        \centering
        \includegraphics[width=\textwidth]{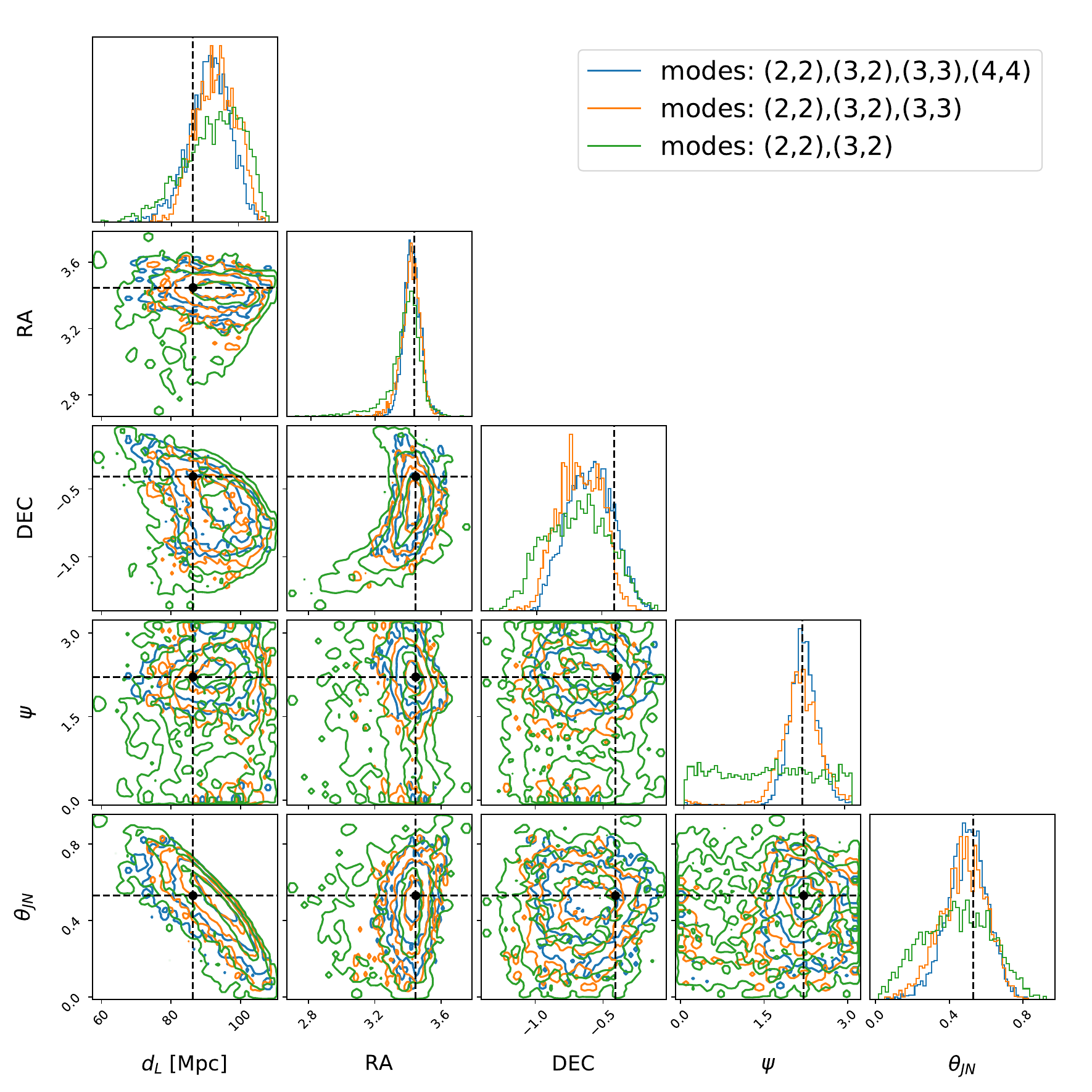}
        \caption{GW1}
        \label{fig:ce_ext2}
    \end{subfigure}
    \hfill 
    \begin{subfigure}{0.49\textwidth}
        \centering
        \includegraphics[width=\textwidth]{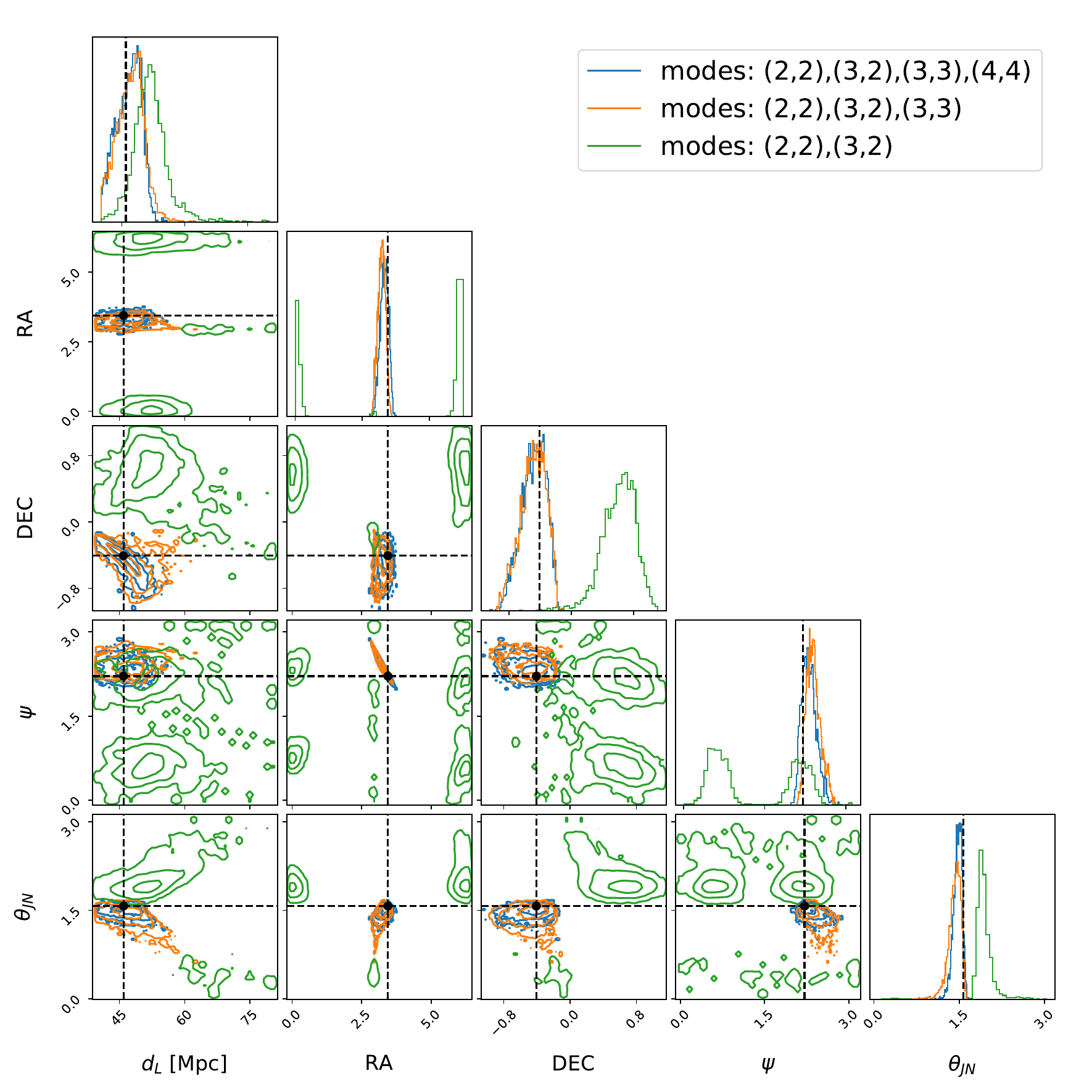}
        \caption{GW2}
        \label{fig:ce_ext1}
    \end{subfigure}
    \caption{Inferred posteriors of extrinsic parameters of GW1 and GW2 in one 40km CE. The green posteriors use all modes for analysis, the orange posterior ignores modes with \texttt{m = 3}, and the blue ones ignore \texttt{m = 3} and \texttt{m = 4} modes. }
    \label{fig:side_by_side_alt}
\end{figure}
We start by looking at the posteriors of GW1 and GW2. For ease of visualization, we break the entire parameter space into intrinsic and extrinsic parameters. 
\subsubsection{Intrinsic Parameters}
The inclusion of higher radiation modes for GW1 has a minimal impact on the intrinsic parameters due to the low signal-to-noise ratios (SNRs) associated with these modes (see Table \ref{table:snr}). In the case of a perfectly face-on orientation ($\theta_{\rm JN} = 0$), the spin-weighted spherical harmonics vanish for modes where \texttt{m} $\neq$ \texttt{2}.  GW1 has an inclination of 0.53, which results in the significant suppression (c.f. table \ref{table:snr}) of the effects of higher-order modes. For GW2, ignoring higher-order modes leads to significant biases in the recovered masses as in figure \ref{fig:ce_int1}.
\subsubsection{Extrinsic Parameters}
For the extrinsic parameters as well, the inclusion of higher modes has a greater impact on GW2 than on GW1. As seen in the case of intrinsic parameters, exclusion of the \texttt{m = 3} mode leaders to biased parameter recovery for GW2 (see figure \ref{fig:ce_ext1}).

The posterior distribution of the polarization angle ($\psi$) presents an interesting case. When higher modes are not included in GW1, the recovered posterior follows the prior distribution (see \ref{fig:ce_ext2}). In contrast, for GW2 without higher modes, the posterior exhibits a bimodal distribution (see \ref{fig:ce_ext1}). A GW with its dominant \texttt{l,m = 2,2} mode for the inspiral portion in a single detector can be written as,
\begin{equation}\label{eq:h}
    h(t) = - \Big(\frac{G\mathcal{M}}{c^2 D_{\text{eff}}}\Big) \Big(\frac{t_0 - t}{5G\mathcal{M}/c^3}\Big)^{1/4}\cos\Big[2 \phi(t-t_0;\mathcal{M},q )+2\phi_c - \tan^{-1}\Big(\frac{F_\times}{F_+}\frac{2 \cos \theta_{\textrm{JN}}}{1+\cos^2 \theta_{\textrm{JN}}}\Big)\Big]  
\end{equation}
Here we only keep terms to the leading order in amplitude. The arrival time at the detector is $t$, while $t_0$ is the arrival time corresponding to the coalescence and
\begin{equation}\label{eq:deff}
    D_\textrm{eff}= D_\textrm{L} \Big[F_+^2 \Big(\frac{1+\cos^2\theta_\textrm{JN}}{2}\Big)^2 + F_\times ^2 \cos^2\theta_\textrm{JN}\Big]^{-1/2}
\end{equation}
For small $\theta_{\texttt{JN}}$ we can write \ref{eq:deff} as,
\begin{equation}\label{eq:deff}
    D_\textrm{eff}= D_\textrm{L} (1-\theta_{\textrm{JN}}^2)^{-1/2}(F_+^2  + F_\times ^2 )^{-1/2} + \mathcal{O}(\theta_{\textrm{JN}}^4)
\end{equation}
If an interferometer has arms along the \textbf{x} and \textbf{y} axes and the source of the gravitational wave is at ($\theta, \phi$), then the beam-pattern function in the long wavelength limit is given by,
\begin{equation}
    F_+ \simeq  -\frac{1}{2}(1+\cos^2 \theta) \cos 2\phi \cos 2\psi - \cos \theta \sin 2\phi \sin 2\psi 
\end{equation}
and 
\begin{equation}
    F_\times \simeq \frac{1}{2}(1+\cos^2 \theta) \cos 2\phi \sin 2\psi - \cos \theta \sin 2\phi \cos 2\psi 
\end{equation}
So,
\begin{equation}
    F_+^2 + F_\times^2 \simeq \frac{1}{4}(1+\cos^2 \theta)^2 + \cos^2 \phi 
\end{equation}
So the amplitude does not depend on the polarization angle to linear order in the inclination.  This is because when the orbit of the binary is a circle in the sky, that circle is invariant under rotations on the plane of the sky. This is true even beyond the long-wavelength limit. The polarization angle enters equation \ref{eq:h} through the antenna responses, and we have shown that the amplitude is independent of $\psi$. The only other place where $\psi$ comes in equation  \ref{eq:h} is via the argument of the cosine term. For small $\theta_{\texttt{JN}}$, the term $\tan^{-1}\Big(\frac{F_\times}{F_+}\frac{2 \cos \theta_{\textrm{JN}}}{1+\cos^2 \theta_{\textrm{JN}}}\Big)$ reduces to $2\psi + f(\theta, \phi)$. So $\phi_c$ and $\psi$ are completely degenerate. Physically, this is because the phase of the binary on the circle in the sky is degenerate with an overall rotation of that circle. For the \texttt{m=2} mode only run, it is not possible to recover any information about $\psi$ unless and until we have higher modes or a network sensitive enough to detect quadratic terms in $\theta_{\rm JN}$. For a face-on system as discussed earlier, the spin-weighted spherical harmonics ($_{-2}Y_{lm}$s) vanish except for \texttt{m=$\pm$2}. So without a \texttt{m=3}, \texttt{m=4} $\phi_c$ and $\psi$ in the waveform's phase would be degenerate \footnote{We do not show the posteriors of $\phi_c$. However, we sample $\phi_c$ for runs containing \texttt{m=3} and \texttt{m=4} modes. For runs containing only the \texttt{m=2} mode, we analytically marginalize over $\phi_c$.}

For GW2, the same cannot be said as it is completely edge-on. In this limit $\tan^{-1}\Big(\frac{F_\times}{F_+}\frac{2 \cos \theta_{\textrm{JN}}}{1+\cos^2 \theta_{\textrm{JN}}}\Big)$ vanishes and the argument of the cosine term in equation \ref{eq:h} is independent of the geometric parameters and there exists no degeneracy with $\phi_c$. The amplitude depends only on $|F_+|$ which is invariant under the transformation $\psi \hookrightarrow \psi + \frac{\pi}{2}$ which is seen in \ref{fig:ce_ext1}. This is a more general degeneracy that exists in the polarization angle \cite{PhysRevD.106.123015} when higher azimuthal modes are ignored.

\subsubsection{Inference with constant antenna response}
\begin{figure}[h!]
    \centering
    \begin{subfigure}{0.49\textwidth}
        \centering
        \includegraphics[width=\textwidth]{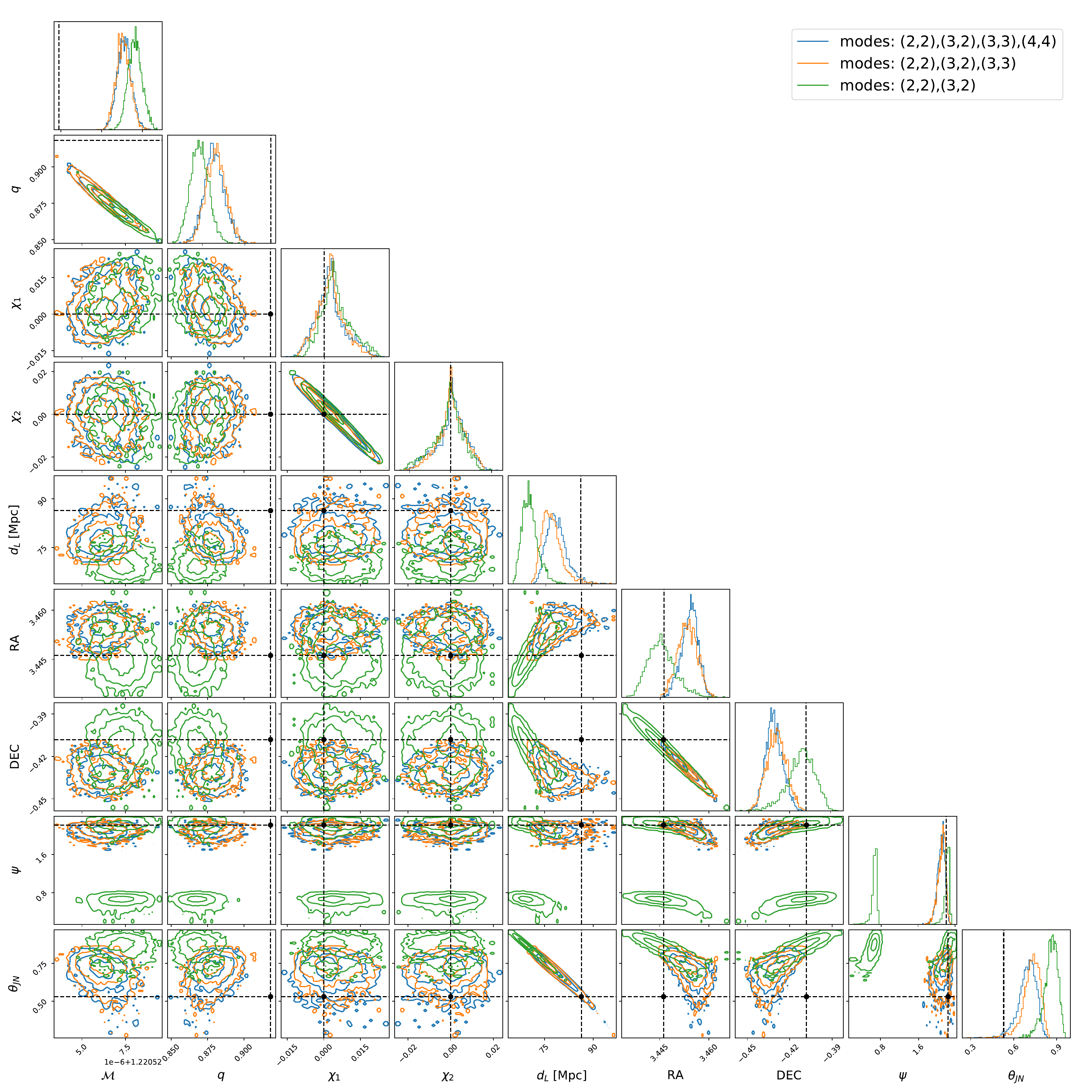}
        \caption{GW1}
        \label{fig:ce_gw1_off}
    \end{subfigure}
    \hfill 
    \begin{subfigure}{0.49\textwidth}
        \centering
        \includegraphics[width=\textwidth]{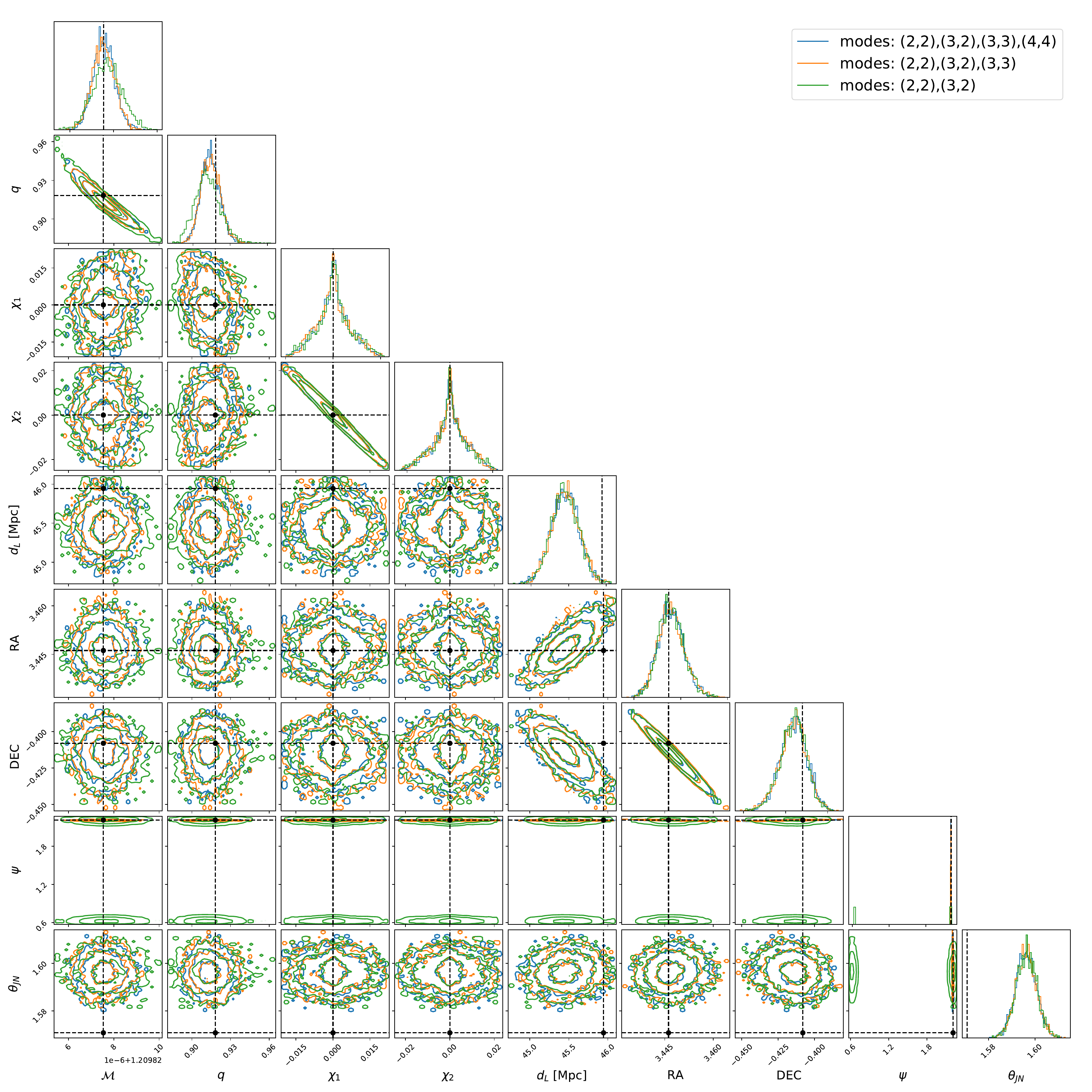}
        \caption{GW2}
        \label{fig:ce_gw2_off}
    \end{subfigure}
    \caption{Inferred posteriors of GW1 and GW2 in one 40km CE with effects due to Earth's rotation and size of the detector turned off.}
    \label{fig:all_off}
\end{figure}

\textcolor{black}{We perform parameter estimation while neglecting Earth's rotation and detector size effects, assuming a static antenna response throughout the signal duration. Specifically, we fix the antenna response to its value at the time of merger. For GW1 (see figure \ref{fig:ce_gw1_off}), where higher-order modes are negligible, we nonetheless observe significant biases in the recovered parameters. This is consistent with \cite{PhysRevD.108.043010}.}

\textcolor{black}{In contrast, for GW2 (see figure \ref{fig:ce_gw2_off}), we recover the intrisic parameters accurately. It is important to note that all simulated signals in this work include contributions from all relevant modes. However, as in the case of GW1, we observe noticeable biases in both the inclination angle and luminosity distance.}

\subsection{Inference using a network of dectectors}
As described earlier, we consider a network of:- CE and A1; CE and CE20; and CE, CE20, and A1. From our simulations, we can draw two conclusions.
\begin{figure}[h!] 
    \centering
    \includegraphics[width=0.8\textwidth]{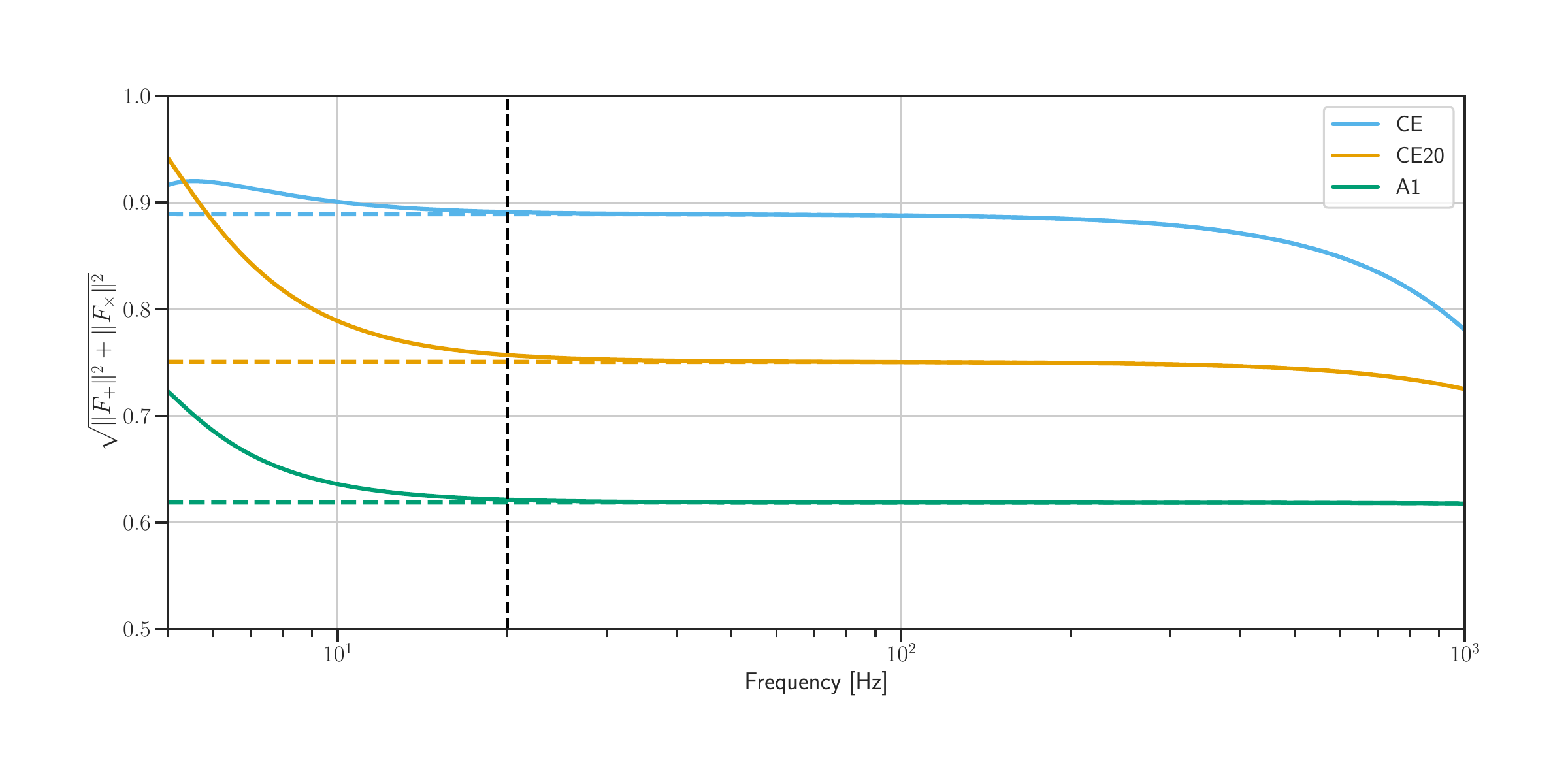}
    \caption{The antenna response for the \texttt{m=2} mode as a function of frequency for GW1 and GW2 at the three detectors is analyzed. For CE and CE20, the response varies at low frequencies due to Earth's rotation and at high frequencies due to detector-size effects. In the case of A1, the signal enters the band at 20 Hz, resulting in no frequency dependence at low frequencies. Additionally, due to the detector's size, there is no effect at high frequencies. \textcolor{black}{The dotted lines denote the antenna response without Earth's rotation effects.}}
    \label{fig:f}
\end{figure}
\begin{itemize}
\item \textbf{Intrinsic parameters}: The network of CE and A1 performs better than only one CE, justifying operating one present-generation detector at design sensitivity with a next-generation detector. The CEA1 network is almost at par with the CECE20 and CECE20A1 networks for GW1 and GW2 as seen in \textcolor{black}{the bottom panels of figure \ref{fig:GW1} and \ref{fig:GW2}}. This is because both these events are of sufficiently high SNR and the phase can be measured accurately in A1. This is not expected to be true at lower SNRs. Also, we do not see much additional benefit from including higher modes of radiation for intrinsic parameters.
\item \textbf{Extrinsic parameters:} Unlike the case for intrinsic parameters, \textcolor{black}{a three-detector network performs better than a two-detector network. This is because three detectors form a larger baseline, breaking degeneracy between the intrinsic parameters. Addition of higher modes does not improve 2D sky localization when we have multiple detectors. However, we see improved luminosity distance, inclination angle and polarization angle recovery (see top panel of \ref{fig:GW1} and \ref{fig:GW2}). For GW1 (see top panel of \ref{fig:GW1}), none of the networks are sensitive to terms of order $\mathcal{O}(\theta_{\rm JN}^4)$ and so we recover the priors on the polarization angle for analysis with only the \texttt{m=2} mode. However, at higher SNRs the networks can become sensitive to the fourth-order term, resulting in a bimodal distribution of the polarization angle. This is because the waveform is invariant under the transformation $\psi \hookrightarrow \psi + \frac{\pi}{2}$, which is a consequence of $|F_+|, |F_\times|$ and $\frac{F_\times}{F_+}$ being invariant under the same transformation. However, higher-order modes are required to choose the correct mode of $\psi$. 
}
\end{itemize}

\begin{figure}[h!]
    \centering
    \begin{subfigure}{0.49\textwidth}
        \centering
        \includegraphics[width=\textwidth]{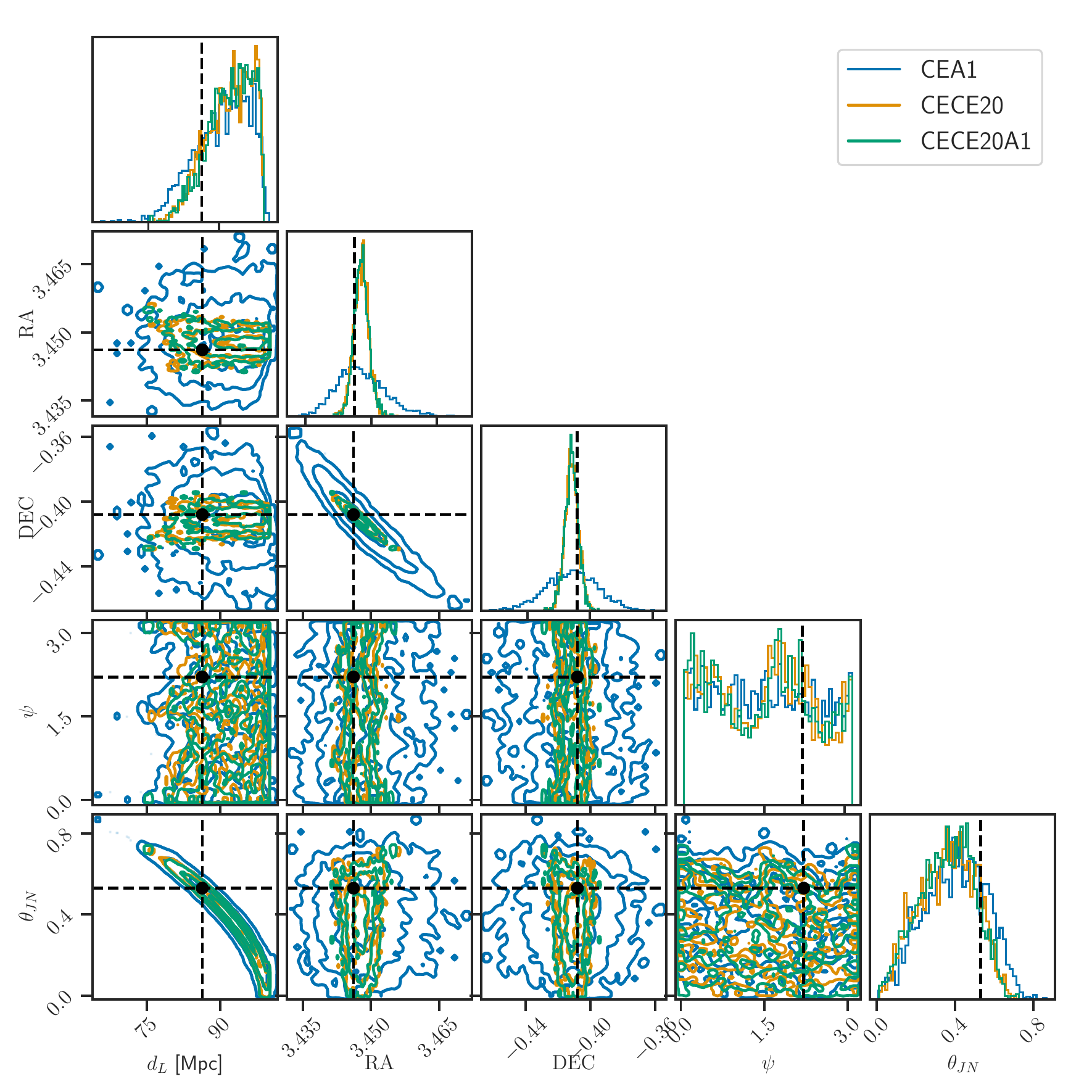}
        \caption{Modes: (2,2),(3,2)}
        \label{fig:ex2_22}
    \end{subfigure}
    \hfill 
    \begin{subfigure}{0.49\textwidth}
        \centering
        \includegraphics[width=\textwidth]{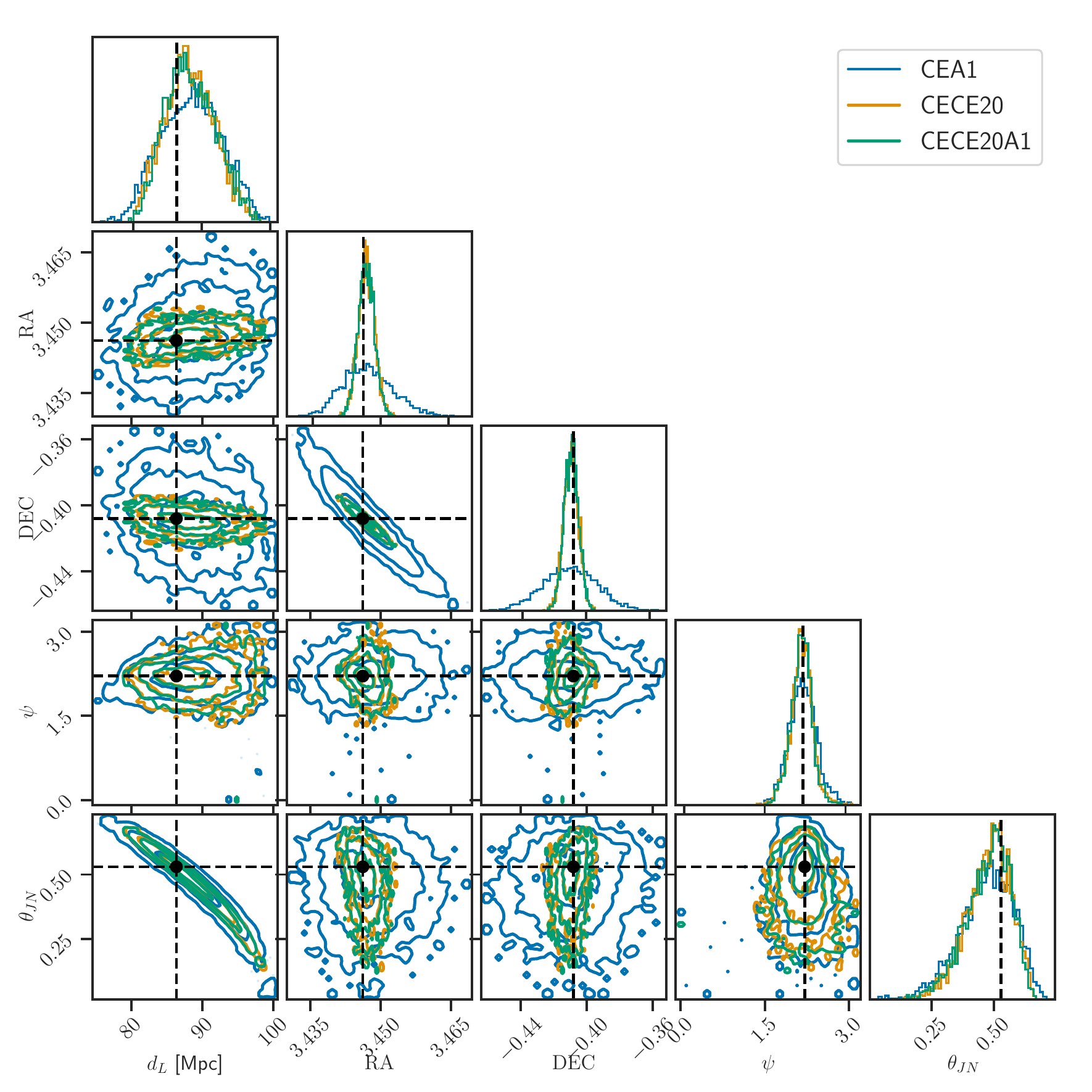}
        \caption{Modes: (2,2), (3,2), (3,3), (4,4)}
        \label{fig:ex2_all}
    \end{subfigure}
    \vskip\baselineskip 
    \begin{subfigure}{0.49\textwidth}
        \centering
        \includegraphics[width=\textwidth]{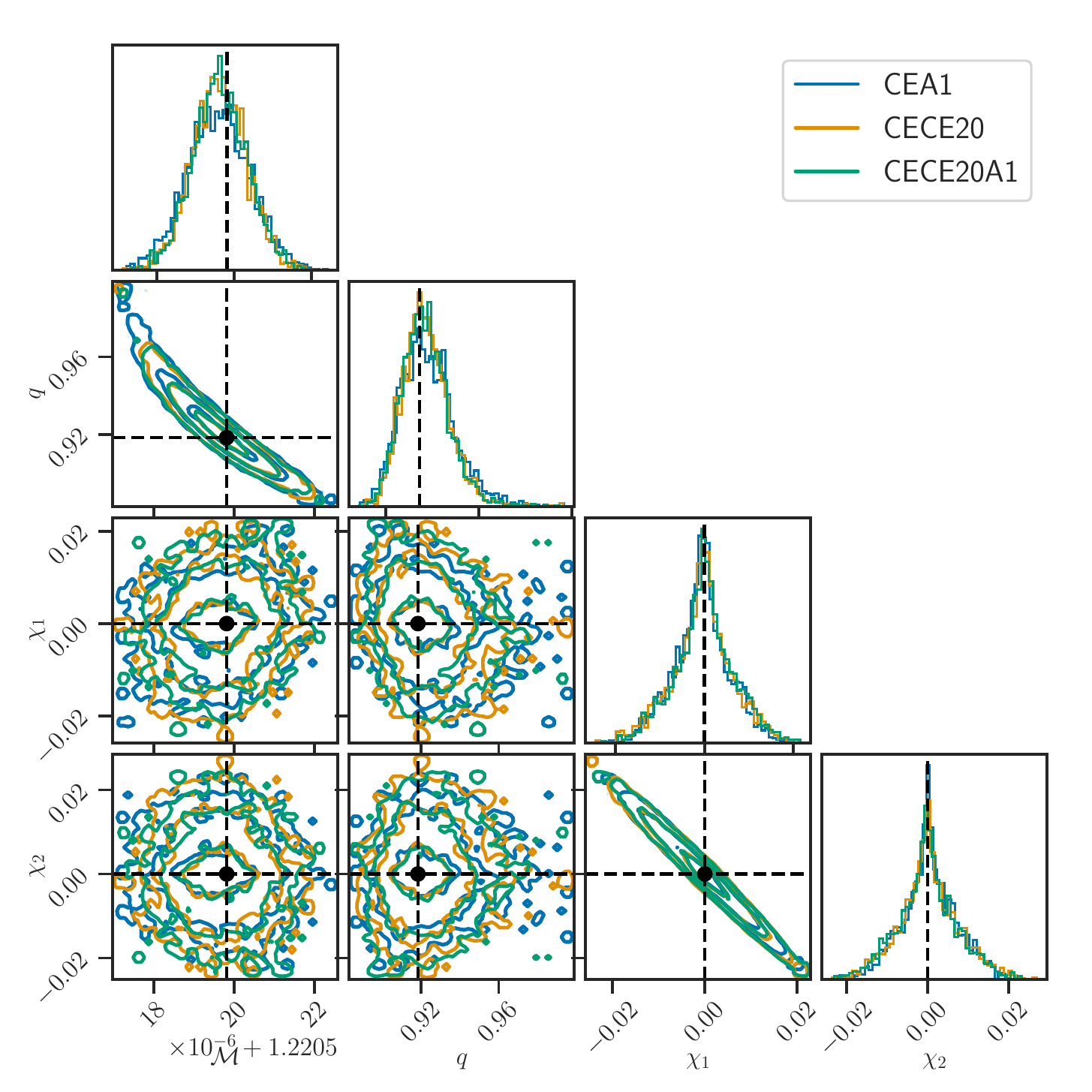}
        \caption{Modes: (2,2), (3,2)}
        \label{fig:in1_22}
    \end{subfigure}
    \hfill 
    \begin{subfigure}{0.49\textwidth}
        \centering
        \includegraphics[width=\textwidth]{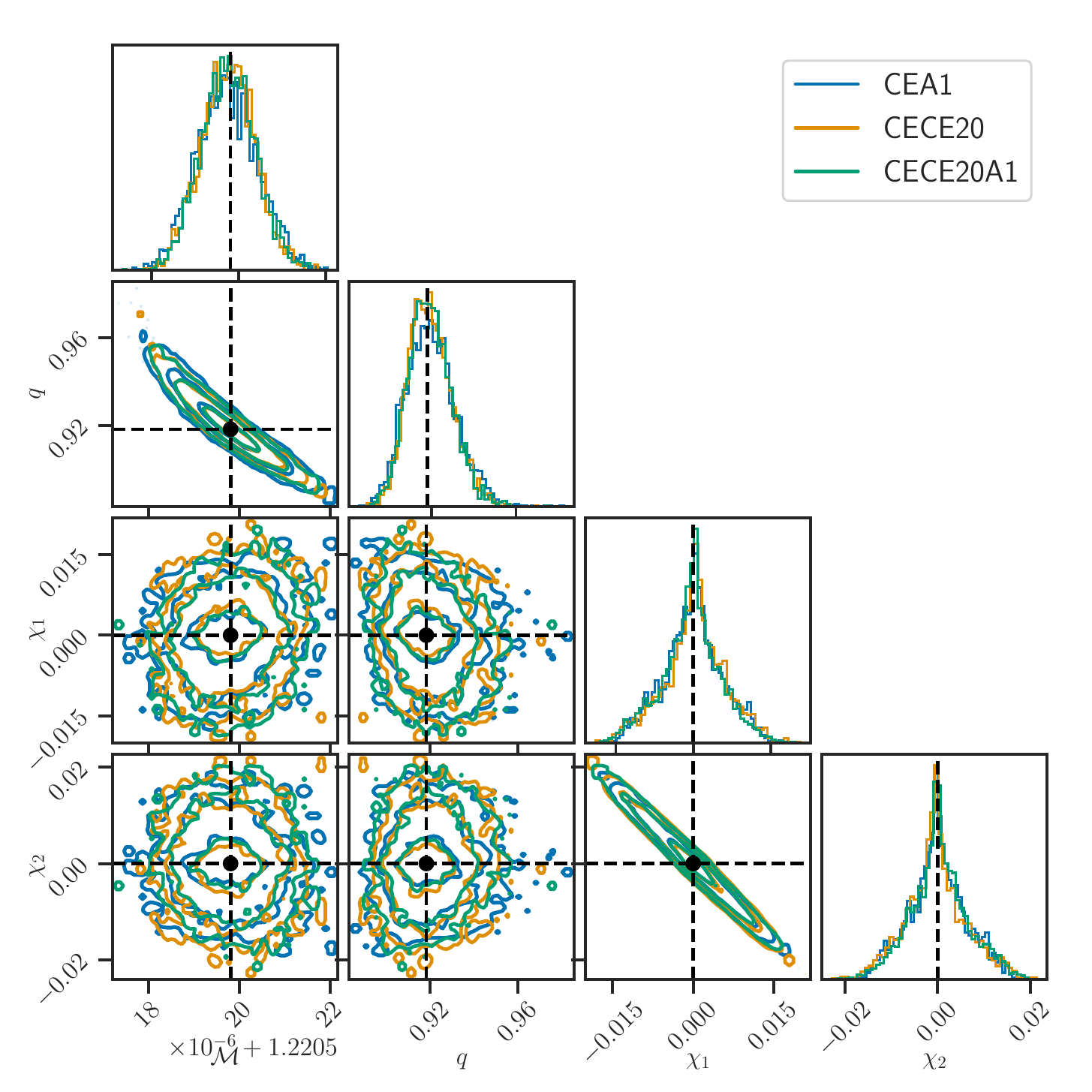}
        \caption{Modes: (2,2), (3,2), (3,3), (4,4)}
        \label{fig:in1_all}
    \end{subfigure}
    \caption{Inferred posteriors of extrinsic parameters \textcolor{black}{(top panel) and intrinsic parameters (bottom panel)} of GW1 in networks comprising of:- a  CE and A1 (blue); CE and CE20 (orange); and CE, CE20 and A1 (green).}
    \label{fig:GW1}
\end{figure}

\begin{figure}[h!]
    \centering
    \begin{subfigure}{0.49\textwidth}
        \centering
        \includegraphics[width=\textwidth]{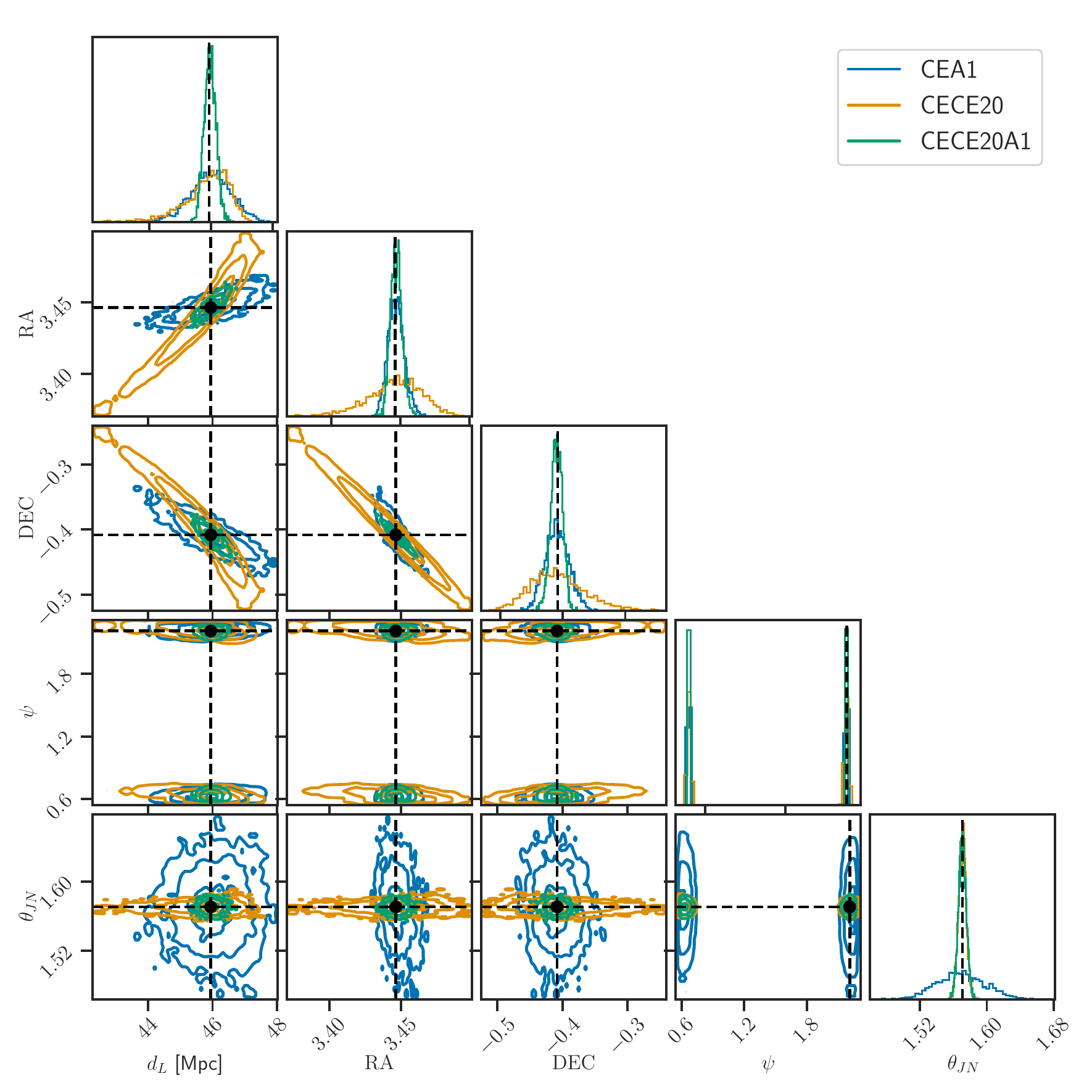}
        \caption{Modes: (2,2), (3,2)}
        \label{fig:ex1_22}
    \end{subfigure}
    \hfill 
    \begin{subfigure}{0.49\textwidth}
        \centering
        \includegraphics[width=\textwidth]{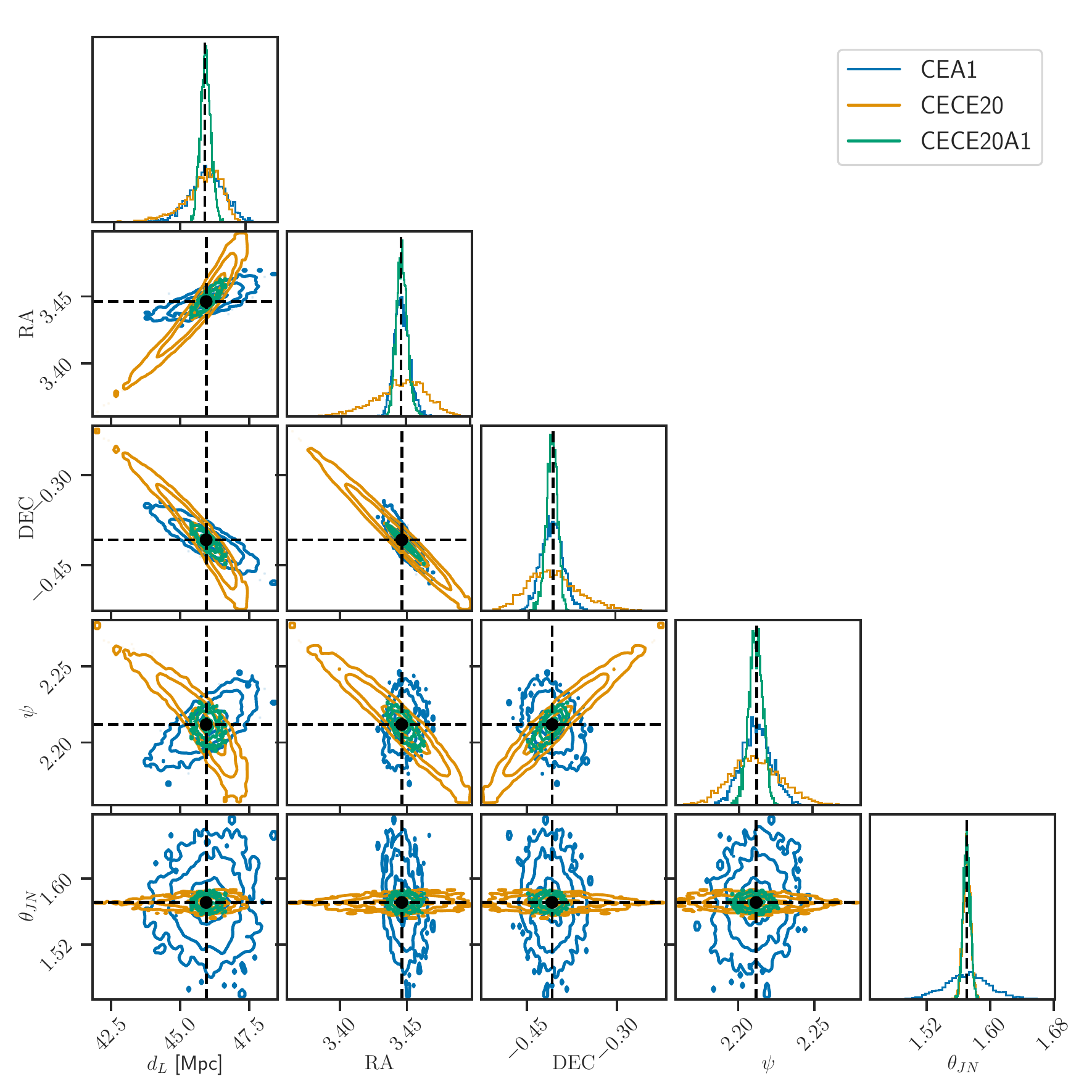}
        \caption{Modes: (2,2), (3,2), (3,3), (4,4)}
        \label{fig:ex1_all}
    \end{subfigure}
    \vskip\baselineskip 
    \begin{subfigure}{0.49\textwidth}
        \centering
        \includegraphics[width=\textwidth]{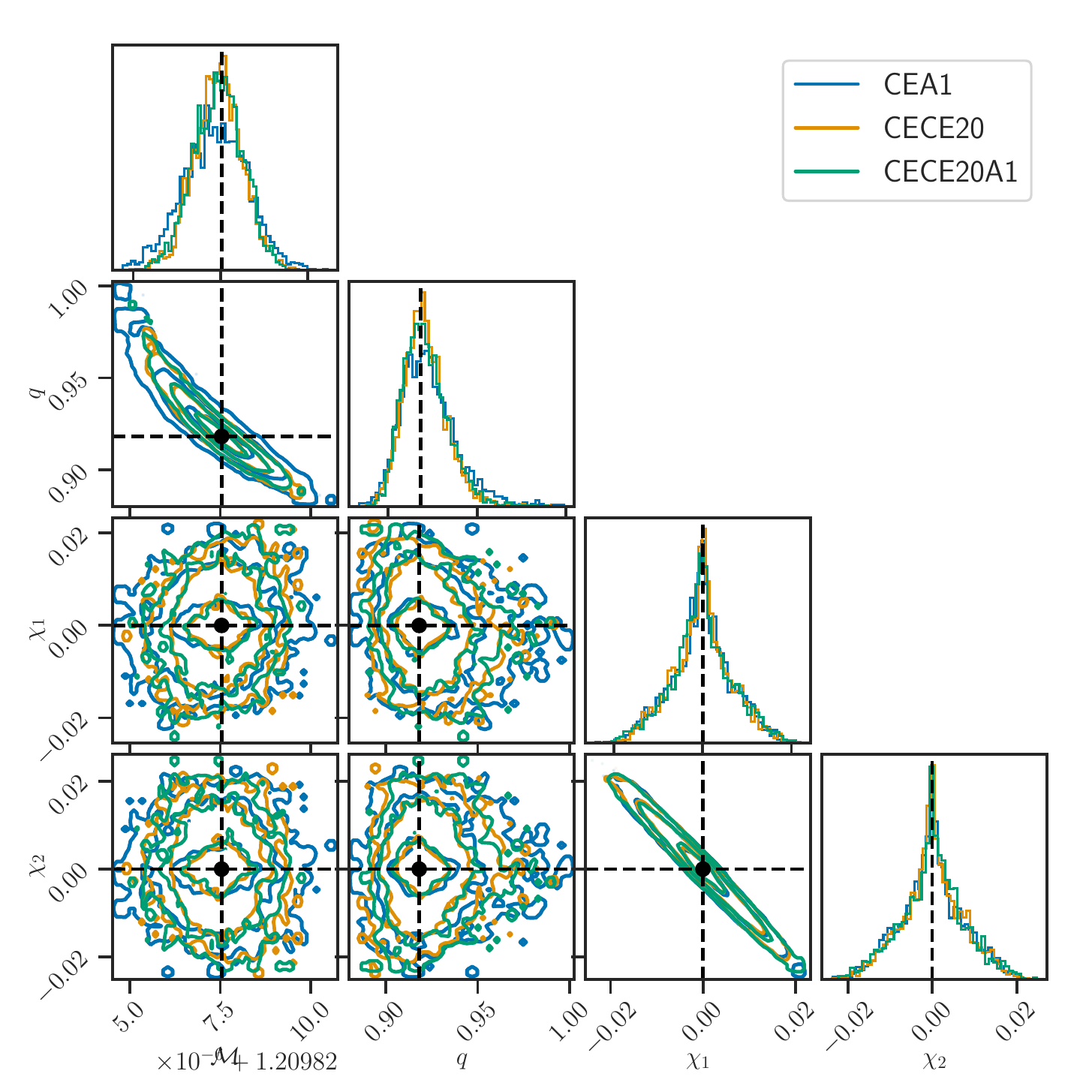}
        \caption{Modes: (2,2), (3,2)}
        \label{fig:in1_22}
    \end{subfigure}
    \hfill 
    \begin{subfigure}{0.49\textwidth}
        \centering
        \includegraphics[width=\textwidth]{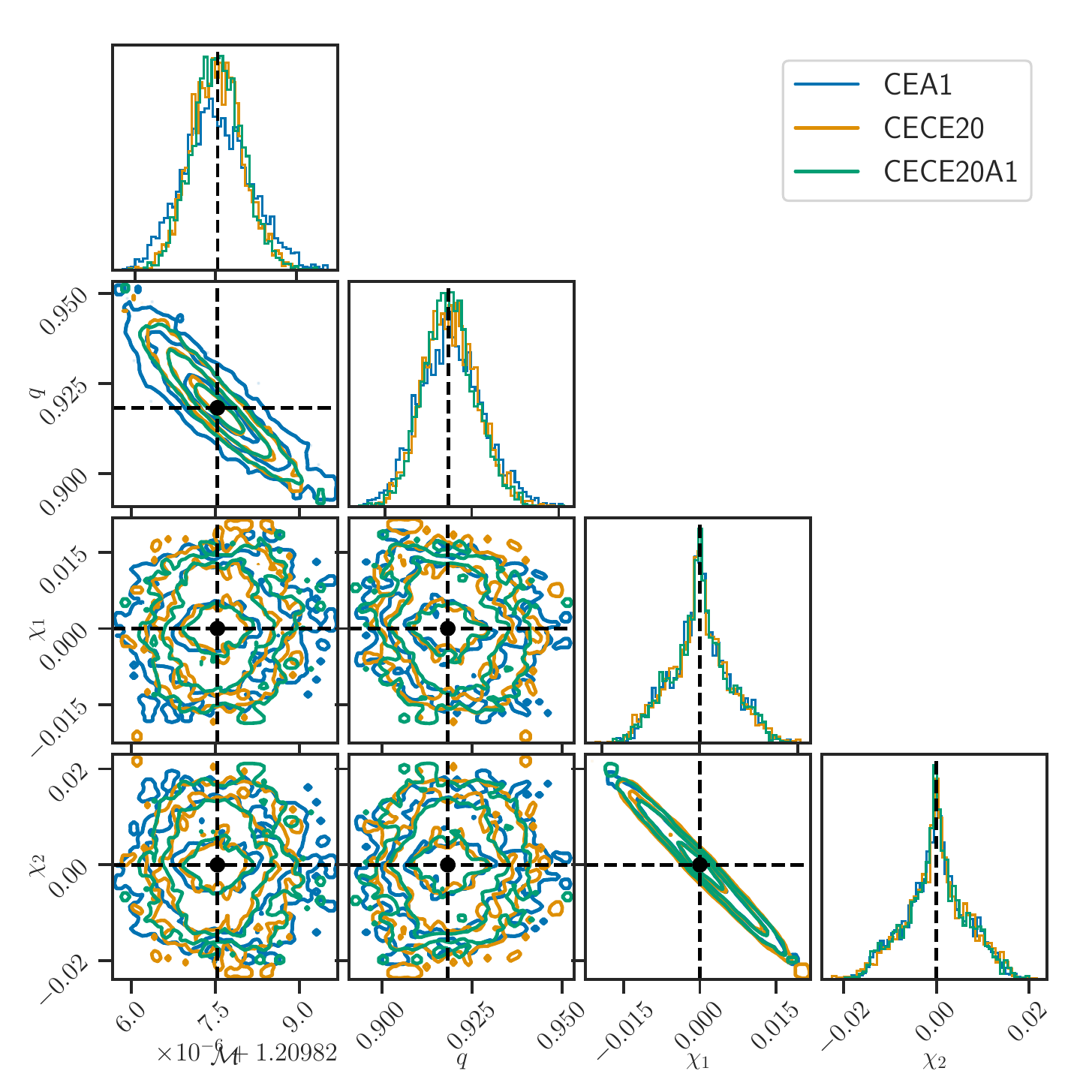}
        \caption{Modes: (2,2), (3,2), (3,3), (4,4)}
        \label{fig:in1_all}
    \end{subfigure}
    \caption{Inferred posteriors of extrinsic (top panel) and intrinsic (bottom panel) parameters of GW2 in networks comprising of: a CE and A1 (blue); CE and CE20 (orange); and CE, CE20 and A1 (green).}
    \label{fig:GW2}
\end{figure}
\subsection{Localization}
Long signals, detector size effects, higher modes, and multiple detectors all contribute to localization. \textcolor{black}{Ignoring effects due to Earth's rotation and detector size results in biased PE and poor sky localization (see figure \ref{fig:all_off}) as the changes in the antenna response (see figure \ref{fig:f}) and the Doppler shift are not properly accounted for}. In the case of only one CE, we observe a 42\% improvement in the 90\% sky area and a 60\% improvement in 3D volume for GW1 (see figure \ref{fig:ce_ext2}) after adding the (3,3) and (4,4) modes. For GW2, ignoring the \texttt{m=3} mode and \texttt{m=4} mode results in biased parameter recovery (see figure \ref{fig:ce_ext1}), as the higher modes of radiation carry significant SNR. The detailed values are given in table \ref{tab:areas}. However, the gain due to higher modes is lost as we add another detector to form a baseline.
\begin{table}[h]
\begin{tabular}{c|cccc|cccc}
\hline
Modes used for inference &
  \multicolumn{4}{c|}{GW1} &
  \multicolumn{4}{c}{GW2} \\ 
 &
  \multicolumn{2}{c|}{\begin{tabular}[c]{@{}c@{}}90\% Sky Area \\ (in sq. deg.)\end{tabular}} &
  \multicolumn{2}{c|}{\begin{tabular}[c]{@{}c@{}}90\% 3D Sky \\ Volume (Mpc$^3$)\end{tabular}} &
  \multicolumn{2}{c|}{\begin{tabular}[c]{@{}c@{}}90\% Sky Area \\ (in sq. deg.)\end{tabular}} &
  \multicolumn{2}{c}{\begin{tabular}[c]{@{}c@{}}90 \% 3D Sky \\ Volume (Mpc$^3$)\end{tabular}} \\ 
 &
  CE &
  \multicolumn{1}{c|}{CEA1} &
  \multicolumn{1}{c}{CE} &
  CEA1 &
  \multicolumn{1}{c}{CE} &
  \multicolumn{1}{c|}{CEA1} &
  \multicolumn{1}{c}{CE} &
  CEA1 \\ \hline
(2,2), (3,2) &
  578 &
  \multicolumn{1}{c|}{\textless 1.5} &
  \multicolumn{1}{c}{51732} &
  77 &
  \multicolumn{1}{c}{-} &
  \multicolumn{1}{c|}{-} &
  \multicolumn{1}{c}{-} &
 - \\ 
(2,2), (3,2), (3,3) &
  313 &
  \multicolumn{1}{c|}{\textless 1.5} &
  \multicolumn{1}{c}{18697} &
  57 &
  \multicolumn{1}{c}{902} &
  \multicolumn{1}{c|}{\textless 3} &
  \multicolumn{1}{c}{4234} &
  \textless 3 \\ 
(2,2), (3,2), (3,3) \& (4,4) &
  \multicolumn{1}{c}{334} &
  \multicolumn{1}{c|}{\textless 1.5} &
  \multicolumn{1}{c}{20373} &
  55 &
  \multicolumn{1}{c}{948} &
  \multicolumn{1}{c|}{\textless 3} &
  \multicolumn{1}{c}{2468} &
  \textless 3 \\ \hline
\end{tabular}
\caption{The 90\% sky areas and 90\% 3d sky volumes for GW1 and GW2 in one 40km CE and a network of CE and A1. For GW2, the run containing only the \texttt{m=2} mode did not recover the position of the injection and hence the sky areas and the sky volumes are not reported. Note that the injections have all modes of radiation.}
\label{tab:areas}
\end{table}

\section{Conclusions}\label{conclusions}
In this paper, we develop the tools necessary to generate GW waveforms for next-generation detectors, incorporating effects due to Earth's rotation, detector size, and higher-order modes, all within the \texttt{Bilby} framework. To handle these waveforms, we implement new likelihood classes, including approximations to the exact likelihood to speed up sampling. We are the first to perform proof-of-concept PE including all these effects.

We simulate gravitational wave signals for next-generation detectors using the \texttt{IMRPhenomXPHM} waveform family. \textcolor{black}{For demonstration purposes, we select a GW170817-like signal, which remains in band for an extended duration, making it well-suited to highlight the significance of Earth-rotation effects. We use black hole waveforms in this analysis, thus setting tidal effects to zero. The waveforms go up to 4 kHz to incorporate detector-size effects (see figure 5) while the $f_{\rm isco}$ for the system is $\sim$ 2kHz. The merger portion of the waveform ($f > f_{\rm isco}$) depends on the neutron star constituents and hence, the high-frequency behavior is unrealistic. This does not cause any biases in our inference, since we are using the same family to generate simulated signals and perform PE. In principle, however, our framework supports the use of BNS waveforms such as \texttt{TEOBResumS} \cite{PhysRevD.98.104052}, \texttt{SEOBNRv5HM} \cite{gamboa2024}, or \texttt{NRHybSur3dq8Tidal} \cite{PhysRevD.102.024031} without modification, aside from the increased computational cost associated with waveform generation. Using a more realistic waveform to perform PE is left for the future.}

For PE, we adopted aligned spin priors and used the computationally efficient mode-by-mode relative binning likelihood to explore the ten-dimensional posterior space. The injected waveform is taken to be the fiducial waveform for simplicity, recognizing that this assumption is infeasible in practical applications. However, this approach does not impact the primary goal of this work, which is to evaluate measurement accuracies in next-generation detectors rather than to develop PE algorithms for future detectors. These PE runs, parallelized across 4 threads on a computing cluster, required approximately one day of wall-clock time. Additional computational optimizations, such as cython-based implementations of beam-pattern functions, could further enhance efficiency, as suggested in the literature \cite{chen2024}.

Baker et al. \cite{baker2025} recently investigated the use of reduced order quadrature (ROQ) models to perform PE on BNS signals while accounting for Earth's rotation and detector size effects. Their study concluded that the computational cost becomes prohibitive when analyzing signals from frequencies below 16 Hz, even when higher-order modes are ignored. However, the problem can be solved using multibanding as demonstrated in \cite{PhysRevD.108.043010}. Our implementation using relative binning enables efficient likelihood evaluation starting from 5 Hz while including higher-order modes, with only negligible loss in accuracy and precision. This demonstrates the scalability and robustness of relative binning for next-generation PE.

While Baker et al. pointed out that multibanding and relative binning may not be sufficiently accurate for analyzing BNS signals with SNRs $\gtrsim 10^3$, we find that the log-likelihood errors introduced by these approximation methods remain below unity even for such high-SNR signals.
A key difference between our study and theirs lies in how the errors are evaluated: we compute them on posterior samples obtained from parameter estimation, whereas they evaluate them on random samples drawn from the prior.
Since relative binning is expected to be accurate primarily in regions where the waveform closely resembles a reference waveform and typically the posterior is concentrated, it is not necessarily accurate across the entire prior volume.
Nevertheless, as demonstrated in our results, parameter estimation using relative binning accurately recovers the injected parameters, indicating that the method is sufficiently accurate in the regions that matter.
For multibanding, we tune the parameters controlling the accuracy to ensure that log-likelihood errors remain below unity, while still achieving a computational speed-up of $\mathcal{O}(10^2)$.

Our approach avoids the costly precomputation of an ROQ basis. Instead, we require only the evaluation of a fiducial waveform, whose parameters are chosen to exactly match those of the injected signal. Our method has virtually zero computational overhead. We note, however, that this idealized setup does not apply to realistic PE, where the true signal parameters are unknown. Generalizing the method to such cases, potentially by iterating on the fiducial waveform or incorporating some hybrid inference schemes using multibanding and relative binning, is to be explored in future studies. We also construct a multibanding likelihood to address the same problem, though we find it computationally expensive to sample from, albeit feasible. As demonstrated in \cite{baker2025}, neglecting Earth's rotation and finite-size detector effects introduces biases in parameter recovery. Nonetheless, for forecasting studies, this configuration provides an accurate and computationally efficient framework for assessing the impact of effects such as Earth's rotation, detector size, and higher-order modes of radiation. It offers a promising foundation for developing scalable inference techniques for next-generation detectors.

We conduct illustrative parameter estimation for two GW170817-like sources: one nearly face-on and the other edge-on. Injections include all modes, while PE runs are performed with and without the subdominant modes. For a single 40 km Cosmic Explorer (CE) detector, the inclusion of HMs improves parameter recovery. However, this advantage diminishes in a network comprising multiple detectors. Regardless, HMs prove essential for accurately measuring the polarization angle and phase, avoiding degeneracies in both cases.

The posteriors we obtain by incorporating all modes of radiation appear fairly Gaussian and lack any multimodalities. Therefore, we anticipate that the Fisher formalism would provide an accurate approximation. However, this remains to be explored in future work.

The accuracy-controlling free parameters of our approximations are calibrated to ensure that errors remain smaller than the statistical uncertainties in PE for SNRs around 1000. For lower SNRs, where statistical uncertainties are naturally larger, the approximation parameters can be adjusted to reduce computational costs further. Additionally, at lower SNRs, the sampling process is easier and quicker. Our current methodology assumes the presence of a single signal in the data; extensions will be necessary to address overlapping signals expected in low-SNR regimes.

We employ the displacement PSD, scaled by the detector length, to ensure that the noise PSD remains free from high-frequency corrections. While this paper primarily focuses on methods for studying CBCs in next-generation detectors, the proposed framework is adaptable to minor variations in PSDs or detector locations, which are yet to be finalized.
 
$Software:$ Analysis in this paper made use of \textsc{Bilby} \cite{bilby_paper}  with new features available in \cite{baral_bilby_xg}, \textsc{LALSuite}v7.2.4 \cite{lalsuite}, \textsc{NumPy}v1.26.4 \cite{Harris2020}, \textsc{SciPy}v1.13.1 \cite{Virtanen2020}, \textsc{Astropy}v6.1.1 \cite{2013, 2018} and \textsc{Matplotlib}v3.7.5 \cite{4160265}.

$Data~and~ code~ availability :$ The codes for this project are hosted in \url{https://git.ligo.org/pratyusava.baral/ce_hm_paper/-/tree/same_snr?ref_type=heads}. The forked version of \textsc{Bilby} containing all the modifications for the next generation detector is available here: \url{https://git.ligo.org/pratyusava.baral/bilby-x-g/-/tree/freq_dep_antenna_response_HM?ref_type=heads}. The data and codes are also available via Zenodo \cite{baral_2025_17087048}.

\section{Acknowledgement}
This work carries LIGO document number P2500078. The authors are thankful to Anson Chen for providing useful insights during the internal LIGO review. This work was supported partially by the National Science Foundation (NSF) awards PHY-2207728 and PHY-2110576 and partially by Wisconsin Space Grant Consortium Awards RFP23\_2-0 and RFP24\_3-0. SM acknowledges support from JSPS Grant-in-Aid for Transformative Research Areas (A) No.~23H04891 and No.~23H04893. The authors are grateful for computational resources provided by the LIGO Laboratory and those provided by Cardiff University, and funded by an STFC grant supporting UK Involvement in the Operation of Advanced LIGO. PB is also grateful to the XG Mock Data Challenge workshop at Penn State University in May 2024.

\bibliographystyle{unsrturl}
\bibliography{main.bib}

\end{document}